\documentclass[aps,twocolumn,nofootinbib]{revtex4}

\usepackage{graphicx}
\usepackage{dcolumn}
\usepackage{bm}

\usepackage{hyperref}
\hypersetup{
    colorlinks=true,
    linkcolor=blue,
    citecolor=blue,
    urlcolor=blue
}

\usepackage{amssymb,amsfonts}
\usepackage{epsfig}
\usepackage{epstopdf}
\usepackage[all]{xy}
\usepackage{amsthm}
\usepackage{dcolumn}
\usepackage{hyperref}
\usepackage{url}
\usepackage{dsfont}
\usepackage{slashed}
\usepackage{mathrsfs}

\usepackage{bbm}
\usepackage{tikz}
\usepackage{physics}
\usepackage{enumitem}
\usepackage{soul}
\usepackage{float}

\newcommand{\be}{\begin{equation}}
\newcommand{\ee}{\end{equation}}
\newcommand{\bea}{\begin{eqnarray}}

\newcommand{\eea}{\end{eqnarray}}

\def\inbar{\,\vrule height1.5ex width.4pt depth0pt}
\def\IR{\relax{\rm I\kern-.18em R}}
\def\IC{\relax\hbox{$\inbar\kern-.3em{\rm C}$}}

\begin{document}

\title{A unified framework for graviton,\\``partially massless'' graviton, and photon fields in de Sitter spacetime\\under conformal symmetry}

\author{Jean-Pierre Gazeau$^{1,2}$\footnote{gazeau@apc.in2p3.fr and j.gazeau@uwb.edu.pl}}

\author{Hamed Pejhan$^{3}$\footnote{pejhan@math.bas.bg}}

\affiliation{$^1$Universit\'e Paris Cit\'{e}, CNRS, Astroparticule et Cosmologie, F-75013 Paris, France}

\affiliation{$^2$Faculty of Mathematics, University of Bia\l ystok, 15-245 Bia\l ystok, Poland}

\affiliation{$^3$Institute of Mathematics and Informatics, Bulgarian Academy of Sciences, Acad. G. Bonchev Str. Bl. 8, 1113, Sofia, Bulgaria}

\date{\today}

\begin{abstract}
We develop a conformally invariant (CI) framework in $(1+3)$-dimensional de Sitter (dS) spacetime, that unifies the descriptions of graviton, ``partially massless'' graviton, and photon fields. This framework is grounded in a rigorous group-theoretical analysis in the Wigner sense and employs Dirac's six-cone formalism. Originally introduced by Dirac, the concept of conformal space and the six-cone formalism were used to derive the field equations for spinor and vector fields in $(1+3)$-dimensional Minkowski spacetime in a manifestly CI form. Our framework extends this approach to dS spacetime, unifying the treatment of massless and ``partially massless'' fields with integer spin $s>0$ under conformal symmetry. This unification enhances the understanding of fundamental aspects of gravitational theories in curved backgrounds.
\end{abstract}

\maketitle

\setcounter{equation}{0} 
\section{Introduction}
This study focuses on examining the CI properties of Wigner massless elementary particles, specifically graviton, ``partially massless'' graviton, and photon fields, in dS spacetime. Let us start from scratch to present the foundational concepts step-by-step in a mathematically rigorous manner.

\subsection{Wigner elementary particles}
In the field theory formulation of elementary particles, Wigner's principles of relativity enforce invariance properties, exemplified by the principle of invariance under the Poincar\'{e} group in flat Minkowski spacetime \cite{Wigner1939, Newton/Wigner, Wigner1952, Levy-Leblond, Voisin, Gursey1963, Fronsdal 1, Fronsdal 2, Aldrovandi, Bacry, Levy, Gazeau2022, Gazeau2023}. 

In the context of flat Minkowski spacetime, quantum elementary particles correspond to (projective) unitary irreducible representations (UIRs) of the Poincar\'{e} group (or one of its coverings) \cite{Wigner1939, Newton/Wigner}. The (rest) mass $m$ and the spin $s$ of an elementary particle serve as the two invariants that characterize the associated UIR of the Poincar\'{e} group \cite{Wigner1939, Newton/Wigner}.

Remarkably, introducing a specific curvature to spacetimes is the sole method for deforming the Poincar\'{e} group. This deformation gives rise to the dS and anti-dS relativity groups of motion \cite{Bacry, Levy, Gazeau2022, Gazeau2023}. This distinctive situation confers a unique status upon dS and anti-dS spacetimes, establishing them as the singular family of curved spacetimes where, to a certain extent, the extension of the notion of elementary particles aligns with Wigner's framework, particularly in connection with UIRs of the relativity group \cite{Gursey1963, Fronsdal 1, Fronsdal 2, Aldrovandi, Bacry, Levy, Gazeau2022, Gazeau2023}. Notably, UIRs of the dS and anti-dS groups, analogous to their common Poincar\'{e} contraction limit, are characterized by two invariant parameters of spin and energy scales.

In this context, and before proceeding further, it is crucial to recall significant advancements in astronomical observations over the past three decades (see, for instance, Refs. \cite{Riess, Perlmutter}) that have yielded a surprising revelation, highlighting the importance of introducing a specific curvature to spacetimes. The present Universe is primarily influenced by a mysterious ``dark'' form of energy density, exhibiting repulsive behavior on a large scale. The most straightforward and widely recognized candidate for this ``dark energy'' is the cosmological constant. In this framework, the dS geometry, serving as the homogeneous and isotropic solution to the cosmological Einstein equations in vacuo, appears to play a dual role. It acts as both the reference geometry for the Universe, representing spacetime devoid of matter and radiation, and the geometry towards which the Universe will asymptotically converge.

\subsection{Wigner massless elementary particles}
Besides Wigner's foundational perspective, it has long been recognized that conformal and gauge invariances are closely tied to the massless nature of elementary particles \cite{Dirac1936, Mack1969, Branson1987, Binegar1983, Mack, Todorov, Barut, Flato, Flato', EEMM, GazeauMurenzi, Gazeaus1, Massless2, Massless2', Massless2'', Bamba1, dSgravity2, dSgravity1}.

In flat Minkowski spacetime, massless elementary particles correspond to UIRs of the Poincar\'{e} group characterized by zero mass and discrete helicity (known as Poincar\'{e} massless representations). These representations uniquely extend to representations of the conformal group \cite{Mack, Todorov}. Specifically, it has been proven that any system invariant under a massless representation of the Poincar\'{e} group is also invariant under a uniquely determined UIR of the conformal group \cite{Mack, Todorov}. Furthermore, in field theories typically described by wave equations, these equations exhibit specific behaviors under conformal transformations in the massless cases \cite{Dirac1936}.

Similarly, massless representations of the dS and anti-dS groups are distinguished as those with a unique extension to the UIRs of the conformal group, \underline{while} that extension is equivalent to the conformal extension of the Poincar\'{e} massless UIRs \cite{Mack, Todorov, Barut, Flato, Flato', EEMM}.

In addition to conformal invariance, the masslessness characteristic of a given Wigner elementary particle (with spin $s>0$) is closely intertwined with a specific gauge-invariant property of the system. This gauge-invariant property, in turn, is associated with the emergence of an indecomposable representation of the respective relativity group, where the corresponding massless UIR (with a unique conformal extension) assumes the central part within the indecomposable representation. For the dS and anti-dS cases relevant to our study, see Refs. \cite{Flato, Flato', EEMM, GazeauMurenzi, Gazeaus1, Massless2, Massless2', Massless2'', Bamba1, dSgravity2, dSgravity1}.

Technically, in the field theory description of Wigner massless elementary particles (with spin $s>0$), the gauge-invariant properties reduce the degrees of freedom from $2s+1$ (as in the respective ``massive'' case) to $2$, specifically the two helicity modes $\pm s$. The propagation of these modes is confined to the light cone. Again, for the dS and anti-dS cases relevant to our study, see Refs. \cite{Flato, Flato', EEMM, GazeauMurenzi, Gazeaus1, Massless2, Massless2', Massless2'', Bamba1, dSgravity2, dSgravity1}.

\subsection{Two specific gauge-invariant cases: dS ``tachyonic'' scalar fields and ``partially massless'' graviton field}
Relevant to the abovementioned context, two specific cases merit careful consideration: the dS ``tachyonic'' scalar fields and the ``partially massless'' graviton field. While both cases possess specific gauge-invariant characteristics, they should not be classified as massless, given that neither has a Minkowskian massless interpretation (in terms of a unique conformal extension, as discussed earlier). Nevertheless, it is entirely valid to examine these fields within a consistent dS framework, both mathematically (via group representation theory) and physically (in terms of field quantization), particularly given the presence of a small but non-zero cosmological constant in real-world cosmology.

The most well-known example among the dS ``tachyonic'' scalar fields is the so-called ``massless'' minimally coupled scalar field, despite lacking any truly massless content! For further discussions on dS ``tachyonic'' scalar fields and ``partially massless'' graviton field, readers are respectively referred to Ref. \cite{dSTachyon} and \cite{PMG} (and \cite{dd1, dd2, dd3, dd4, dd5, dd6}), and references therein.

Nevertheless, despite lacking genuine massless content, the ``partially massless'' graviton field possesses the intriguing feature of light-cone propagation, making it relevant to the present study, as conformal invariance and light-cone propagation are closely intertwined. It exhibits specific gauge invariance, reducing its degrees of freedom to $4$, which is less than the $5$ of a ``massive'' graviton field but more than the two helicities $\pm 2$ of the graviton (strictly massless spin-$2$) field \cite{PMG}. The corresponding physical degrees of freedom propagate on the light cone, a property that gives rise to the term ``partially massless'' \cite{PMG}.

\subsection{Objectives and layout}
This paper explores the conformal properties of the dS graviton field, utilizing the concepts of conformal space and Dirac's six-cone formalism. It extends the framework developed by Gazeau et al. (1989) for anti-dS spacetime \cite{GazeauHans}. Our analysis uncovers a remarkable structure that unifies the graviton (strictly massless spin-$2$), the ``partially massless'' graviton, and massless vector (photon) fields within the framework of conformal symmetry.

The remainder of this manuscript is structured as follows to achieve our objectives. Sect. \ref{Sec. Machinery} provides a brief introduction to the dS machinery and Dirac's six-cone formalism. In Sect. \ref{Sec. III}, we consider a symmetric rank-$2$ tensor field on the five-dimensional surface of Dirac's six-cone in $\mathbb{R}^6$ and derive the possible reduced fields in dS spacetime. This approach consistently leads to the graviton (strictly massless spin-$2$), the ``partially massless'' graviton, and the massless vector (photon) fields in dS spacetime. Finally, we discuss our findings in Sect. \ref{Sec. conclusion}. This paper is supplemented with four appendices, in which the mathematical details of the calculations are provided.

\subsection{Convention}
Throughout this article, we use units where $\boldsymbol{c} = 1$ and $\boldsymbol{\hbar} = 1$, with $\boldsymbol{c}$ representing the speed of light and $\boldsymbol{\hbar}$ representing the Planck constant.

\section{dS machinery and Dirac's six-cone formalism}\label{Sec. Machinery}

\subsection{dS machinery}
The $(1+3)$-dimensional dS manifold is topologically $\mathbb{R}^1 \times \mathbb{S}^3$, where $\mathbb{R}$ denotes a timelike direction. This manifold can be depicted as a hyperboloid embedded in a $(1+4)$-dimensional Minkowski spacetime $\mathbb{R}^5$:
\begin{align}
    \text{dS} = \left\{ x = (x^a)\in \mathbb{R}^5 \;;\; (x)^2 \equiv x\cdot x = \eta^{}_{ab} x_{}^a x_{}^b = - R^2 \right\}\,,
\end{align}
where $a,b = 0,1,2,3,4$, $\eta^{}_{ab} = \mbox{diag}(1,-1,-1,-1,-1)$ defines the $(1+4)$ metric of $\mathbb{R}^5$, and $R$ represents a positive constant denoting the radius of the dS hyperboloid. 

The dS (relativity) group is SO$_0(1,4)$ — indicating the connected subgroup of O$(1,4)$ — or its universal-covering group Sp$(2,2)$. The associated Lie algebra is characterized by the linear span of the (ten) Killing vectors:
\begin{align}
    K_{ab} = x^{}_a\partial^{}_b - x^{}_b\partial^{}_a \,.
\end{align}

On the representation level, in the Hilbert space of the symmetric, homogeneous, transverse, \emph{square-integrable}\footnote{With reference to an invariant inner product of the Klein-Gordon type or a similar form.} rank-$r$ tensor fields ${\psi}^{(r)}_{a_1 \,\ldots\, a_r}$ defined on the dS hyperboloid,\footnote{Note that only integer spin cases are discussed here.} the Killing vectors $K_{ab}$ are represented by (essentially) self-adjoint operators $L^{(r)}_{ab} = M^{}_{ab} + S^{(r)}_{ab}$, where the orbital part is given by \cite{Moylan}:
\begin{align}
    M_{ab} = - \mathrm{i} \left( x^{}_a\partial^{}_b - x^{}_b\partial^{}_a \right)\,, 
\end{align}
and the action of the spinorial part on the tensorial indices is given by:
\begin{align}
    S_{ab}^{(r)}\, {\psi}^{(r)}_{a_1 \,\ldots\, a_r} = - \mathrm{i} \sum_{i=1}^{r}\Big(\eta^{}_{aa_i} {\psi}^{(r)}_{a_1 \,\ldots\, (a_i\mapsto b) \,\ldots\, a_r} - (a \rightleftharpoons b)\Big)\,.
\end{align}
The representative operators $L^{(r)}_{ab}$ verify the commutation rules of the dS Lie algebra:
\begin{align}\label{adjoint commutator}
\left[L^{(r)}_{ab},L^{(r)}_{cd}\right] =  \mathrm{i} \left( \eta^{}_{ac} {L^{(r)}_{bd}} + \eta^{}_{bd} {L^{(r)}_{ac}} - \eta^{}_{ad} {L^{(r)}_{bc}} - \eta^{}_{bc} {L^{(r)}_{ad}} \right)\,.
\end{align}
They also verify the commutation relation:
\begin{align}
    \left[L^{(r)}_{ab},(x)^2\right]=0\,,
\end{align}
which implies that $L^{(r)}_{ab}$s are intrinsically defined on the dS hyperboloid $(x)^2 = -R^{2}$.

In this context, two Casimir operators emerge:
\begin{align}
\label{Casimir 2} Q^{(1)}_r =&\, - \frac{1}{2} L^{(r)}_{ab} L^{(r)ab} \quad (\text{quadratic})\,, \\
Q^{(2)}_r =&\, - W^{(r)}_a W^{(r)a} \quad (\text{quartic})\,,
\end{align}
where $W^{(r)}_a = - \frac{1}{8} {\cal{E}}_{\tiny{abcde}} L^{(r)bc} L^{(r)de}$ stands for the counterpart of the Pauli-Lubanski operator in the dS context (${\cal{E}}_{\tiny{abcde}}$ is the five-dimensional totally antisymmetric Levi-Civita symbol). These Casimir operators are also intrinsically defined on the dS hyperboloid, where they satisfy the commutation relation:
\begin{align}
    \left[Q^{(1,2)}_r,(x)^2\right]=0\,.
\end{align}
Moreover, they obey the following commutation rules:
\begin{align}
    \left[Q^{(1,2)}_r, L^{(r)}_{ab}\right]=0\,, \quad \forall a,b = 0,1,\ldots,4\,.
\end{align}

Given the latter equation, the Casimir operators act as constants on all states ${\psi}^{(r)}_{a_1 \,\ldots\, a_r}$ within a certain dS UIR:
\begin{align}\label{wa.eq}
    Q^{(1,2)}_r \, {\psi}^{(r)}_{a_1 \,\ldots\, a_r} = \langle Q^{(1,2)}_r \rangle \, {\psi}^{(r)}_{a_1 \,\ldots\, a_r}\,,
\end{align}
where $\langle Q^{(1,2)}_r \rangle$ stands for the respective eigenvalues. According Dixmier \cite{Dixmier}, the eigenvalues $\langle Q^{(1,2)}_r \rangle$ are characterized by a pair of parameters $(p\in \mathbb{N}/2 \,,\, q\in\mathbb{C})$ as follows:
\begin{align}
    \label{Casimir rank 2} \langle Q^{(1)}_r \rangle =&\, \big(-p(p+1) - (q+1)(q-2)\big)\,, \\
    \label{Casimir rank 4} \langle Q^{(2)}_r \rangle =&\, \big(-p(p+1)q(q-1)\big)\,.
\end{align}

Considering the eigenvalue equations \eqref{wa.eq}, the spectral values of the Casimir operators, specifically the allowed values of $(p \,,\, q)$, serve to classify UIRs of the dS group. These UIRs are generally categorized into three distinct series \cite{Gazeau2022, Thomas, Newton, Takahashi, Dixmier, Martin1974}: 
\begin{enumerate}
    \item{The principal series, characterized by:
          \begin{enumerate}
              \item{$p=s=0,1,2, \ldots$ and $q = 1/2 \pm \mathrm{i}\nu$, with $\nu \in \mathbb{R}$,}
              \item{$p=s=1/2,3/2, \ldots$ and $q = 1/2 \pm \mathrm{i}\nu$, with $\nu \in \mathbb{R} - \{0\}$.}
          \end{enumerate}
          Note that: (i) $p=s$ corresponds to the spin interpretation. (ii) The two sets of representations labeled by $\nu$ and $-\nu$ are equivalent.}
    \item{The complementary series, characterized by:
          \begin{enumerate}
              \item{$p=s=0$ and $q = 1/2 \pm \nu$, with $\nu \in \mathbb{R}$ and $0<|\nu|<3/2$,}
              \item{$p=s=1,2, \ldots$ and $q = 1/2 \pm \nu$, with $\nu \in \mathbb{R}$ and $0<|\nu|<1/2$.}
          \end{enumerate}
          Note that: (i) $p=s$ has a spin meaning. (ii) The representations corresponding to $\nu$ and $-\nu$ are equivalent.}
    \item{The discrete series, characterized by:
          \begin{enumerate}
              \item{$p=1,2, \ldots$ and $q=0$,}
              \item{$p=1/2,1,3/2, \ldots$ and $q=p,p-1, \ldots , 1$ or $1/2$.}
          \end{enumerate}
          Note that $p=s=q$ has a spin (helicity) meaning.}
\end{enumerate}

In the null-curvature limit, the dS principal series UIRs contract to the massive UIRs of the Poincar\'{e} group, comprehensively covering the entire set of the latter and earning the designation of dS massive representations \cite{Mickelsson, Garidi}. While no dS UIR is comparable to the Poincar\'{e} massless infinite-spin UIRs, a specific member of the dS scalar complementary series UIRs, i.e., $(p=s=0 \,,\, q=1)$, along with the dS higher-spin discrete series UIRs at the lower end of this series, i.e., $(p=s \,,\, q=s)$, form a distinctive set of the dS UIRs with a unique extension to the conformal group (SO$_0(2,4)$) UIRs \cite{Barut, Mack}. This extension precisely aligns with the conformal extension of the massless UIRs of the Poincar\'{e} group \cite{Barut, Mack}. Consequently, these representations are referred to as the dS massless UIRs.\footnote{Notably, this correspondence also holds at the level of the field equations.} All other dS UIRs, not falling strictly into the massive or massless categories, either possess a nonphysical Poincar\'{e} contraction limit or lack such a limit altogether.

A key insight here is that, in practical terms, for a given dS UIR, the common dense subspace within its corresponding Hilbert space — the carrier space of the UIR — consists of \emph{square-integrable}\footnote{With respect to some invariant inner product, such as the Klein-Gordon.} eigenfunctions ${\psi}^{(r)}_{a_1 ,\ldots, a_r}$ of the eigenvalue equations \eqref{wa.eq}, specifically tailored to the associated Casimir eigenvalues \cite{Gazeau2022}. In the quantum field theory (QFT) description of the corresponding elementary particle, the equation involving the quartic Casimir operator $Q^{(2)}_r$, which includes higher derivatives, naturally entails ``ghost'' solutions. Consequently, the equation for the quadratic Casimir operator $Q^{(1)}_r$ assumes a fundamental role as the respective ``field (wave) equation'' in this group-theoretical framework:
\begin{align}\label{fieldEq}
   \left( Q^{(1)}_r - \langle Q^{(1)}_r \rangle \right) {\psi}^{(r)}_{a_1 \,\ldots\, a_r} = 0\,,
\end{align}
with the specific allowable values of the parameters $p$ and $q$ corresponding to the three series of dS UIRs \cite{Gazeau2022}.

Our objectives in this study necessitate a deeper exploration of the aforementioned field equation for rank-$r=0,1$, and $2$ tensor fields within the context of ambient space notation. Let us begin with the simplest case: a rank-$0$ tensor (scalar) field.

\emph{\textbf{Remark:} From this point onward, we simplify the notation by denoting the quadratic Casimir operator as $Q_r^{(1)} \equiv Q_r$.}

\subsubsection{Rank-$0$ tensor (scalar) field}\label{Sect. scalar}
We denote a dS scalar field as $\phi \equiv {\psi}^{(0)}_{}$. It is supposed to be homogeneous with, for simplicity, the degree of homogeneity zero, meaning $x\cdot\partial \phi = 0$. Additionally, the field is considered square-integrable with respect to the Klein-Gordon inner product. The quadratic (scalar) Casimir operator $Q_0$, applied to the space spanned by $\phi$, reads as \cite{Gazeau2022}:
\begin{align}\label{Q_0}
    Q_0 = -\frac{1}{2} M_{ab}M^{ab}= -R^{2} \bar\partial^2\,,
\end{align}
where $\bar\partial_a \equiv \theta_{ab} \partial^b = \partial_a + R^{-2} x^{}_a x\cdot \partial$ is known as the transverse derivative, and $\theta_{ab} \equiv \eta^{}_{ab} + R^{-2} x^{}_a x^{}_b$ is the transverse projector. Notably, $x\cdot\theta \equiv x^a_{}\theta_{ab} = x^b_{}\theta_{ab} = 0$, and hence $x\cdot\bar\partial = 0$. Two significant identities arise here: 
\begin{align}
    \bar\partial^{}_a x^{}_b = \theta_{ab}\,, \quad \bar\partial^{}_a (x)^2 = 0\,.
\end{align}
The latter implies that the transitive operator $\bar\partial$ commutes with $(x)^2$, highlighting its intrinsically defined nature on the dS manifold.

In this context, the field equation \eqref{fieldEq} reduces to:
\begin{align}
   \left( Q_0 - \langle Q_0 \rangle \right) \phi = \left( Q_0 + \tau (\tau + 3) \right) \phi = 0\,,
\end{align}
where $\tau = -q-1$ for the principal and complementary series, and $\tau = -p-2$ for the discrete series, with $p$ and $q$ in the allowed parameter ranges associated with the three series of the dS UIRs. In the specific case of the scalar massless elementary particle associated with the scalar complementary UIR $(p = s = 0 \,,\, q = 1)$, which is of particular interest in this study, the scalar field equation reduces to:
\begin{align}
   \left( Q_0 - 2 \right) \phi = 0\,.
\end{align}
The corresponding field is known in the literature as the dS conformally coupled massless scalar field.

\subsubsection{Rank-$1$ tensor (vector) field}\label{Subsect. 1}
Now, let us focus on the specific case of a vector field $K_a \equiv {\psi}^{(1)}_{a_1}$,\footnote{From now on, whenever possible, we omit tensorial indices for simplicity.} within the framework of ambient space notation. By definition, $K$ is a homogeneous, transverse, and \emph{square-integrable}\footnote{Under some invariant Klein-Gordon type inner product.} vector field on the dS manifold. In this notation, the quadratic Casimir operator $Q_1$, acting on the space spanned by $K_a$, is typically represented as \cite{Gazeau2022}:
\begin{align}\label{Q1vec}
Q_1 K_a = \left( Q_0 - 2 \right)K_a - \underbrace{2 \partial_a x\cdot K}_{=0} + 2 x^{}_a \partial\cdot K \,.
\end{align}
Note that the second term on the right-hand side vanishes due to the transversality condition intrinsic to fields in ambient space notation.

Considering the structure of $Q_1$ in ambient space notation, the solution space of the associated field equation for $K$ includes invariant subspaces that must be excluded to isolate the space solely hosting the relevant dS UIR. Therefore, in addition to the conditions inherent to fields in ambient space notation, we further impose the auxiliary condition of divergencelessness $\partial \cdot K = 0$.\footnote{Note that for any transverse dS field, such as $K$, the relation $\partial\cdot K = \bar\partial\cdot K$ holds.} Consequently, the field equation \eqref{fieldEq} for a given vector field $K$ ($x \cdot K = 0 = \partial \cdot K$) simplifies to:
\begin{align}\label{Field Eq. vec}
&\left( Q_1 - \langle Q_1 \rangle \right) K \nonumber\\
& = \left( Q_0 - 2 + p(p+1) + (q+1)(q-2) \right) K = 0\,,
\end{align}
with the specific allowable values of the parameters $p$ and $q$ corresponding to the three series of dS UIRs.

The crucial observation in this context is that the above field equation exclusively captures the \underline{physical} degrees of freedom, associated with a specific UIR of the dS group, for the dS vector field under consideration. Therefore, in theories lacking gauge invariance, this equation comprehensively encompasses all the degrees of freedom. However, when examining the gauge invariance properties of the dS massless vector field associated with the discrete UIR $(p=1 \,,\, q=1)$, or equivalently associated with the quadratic Casimir eigenvalue $\langle Q_1 \rangle = 0$, the equation proves to be overly restrictive, as it fails to include the corresponding gauge degrees of freedom. 

Technically, in the massless case, the solution space for $Q_1 K = 0$ exhibits a singularity of the form $1/\langle Q_1 \rangle$ (see Ref. \cite{Gazeaus1}). To eliminate this singularity, the condition of divergencelessness on $K$ ($\partial \cdot K = 0$), necessary for associating $K$ with the aforementioned massless UIR of the dS group, must be relaxed \cite{Gazeaus1}; the field equation $Q_1 K = 0$ must be solved in an expanded solution space that includes configurations where $\partial \cdot K \neq 0$. As a result, the equation is replaced by the following gauge-invariant form:
\begin{align}
    Q_1 K + D_1 \partial \cdot K = 0 \,,
\end{align}
where $D_1= R^{2}\bar\partial$, such that:
\begin{align}
    K \mapsto K + D_1 \phi_g
\end{align}
is a solution to the field equation for any scalar field $\phi_g$, provided $K$ is. Then, there are three main types of solutions for $K$: pure gauge solutions, physical solutions that are divergenceless, and non-divergenceless solutions. The full solution space forms an indecomposable representation of the dS group, with the massless UIR \( (p = 1 \,,\, q = 1) \) as its central, physical component. Notably, the space of one-particle physical states is associated with this massless UIR. For the explicit structure of this indecomposable representation, see Ref. \cite{Gazeaus1}. By introducing a gauge-fixing parameter, denoted as $c$, the explicit form of the field equation reads as \cite{Gazeaus1}:
\begin{align}\label{masslessvector}
    Q_1 K + cD_1 \partial \cdot K = 0 \,.
\end{align}

\subsubsection{Rank-$2$ tensor field}\label{Subsect. 2}
By definition, in ambient space notation, a rank-$2$ tensor field ${\cal{K}}_{ab} \equiv {\psi}^{(2)}_{a_1a_2}$ is a symmetric (${\cal{K}}_{ab} = {\cal{K}}_{ba}$), homogeneous, transverse, and square-integrable entity defined on the dS hyperboloid.\footnote{Again, whenever possible, we omit tensorial indices for simplicity.} In this notation, the quadratic Casimir operator $Q_2$, applied to the space spanned by ${\cal{K}}_{ab}$, is generally expressed as \cite{Gazeau2022}:
\begin{align}\label{Q1ten}
Q_2 \, {\cal{K}}_{ab} =&\, \left( Q_0 - 6 \right) {\cal{K}}_{ab} - \underbrace{2 {\cal{S}} \partial_a x\cdot {\cal{K}}_{\cdot b}}_{=0} + 2 {\cal{S}} x^{}_a \partial\cdot {\cal{K}}_{\cdot b} \nonumber\\
& + 2 \eta_{ab} {\cal{K}}^\prime \,,
\end{align}
where ${\cal S}$ denotes the symmetrizer operator (${\cal S}(\zeta_a \omega_b)=\zeta_a \omega_b + \zeta_b \omega_a$) and ${\cal{K}}^\prime \equiv \eta^{ab} {\cal{K}}_{ab}$ represents the trace of the field. Similar to the vector case, the second term on the right-hand side vanishes due to the transversality condition inherent in the fields described by ambient space notation.

Again, considering the structure of $Q_2$ in ambient space notation, the solution space of the associated field equation for ${\cal{K}}$ includes invariant subspaces that need to be excluded to isolate the space solely hosting the relevant dS UIR. Hence, besides the conditions inherent to the fields in ambient space notation, we further impose the auxiliary condition of divergencelessness ($\partial \cdot {\cal{K}} = 0$); it is important to note that the combined requirements of transversality and divergencelessness imply ${\cal{K}}^{\prime} = 0$.\footnote{To illustrate the point, we start from $\partial^a {\cal{K}}_{ab} = 0$. By contracting the ``free'' index, we naturally have:
\begin{align*}
    -x^b (\partial^a {\cal{K}}_{ab}) = 0 \; \Rightarrow \; \underbrace{-\partial^a x^b {\cal{K}}_{ab}}_{=0} + \eta^{ab}_{} {\cal{K}}_{ab} = {\cal{K}}^\prime = 0\,.
\end{align*}} Subsequently, the field equation \eqref{fieldEq} for a given rank-$2$ tensor field ${\cal{K}}$ reduces to:
\begin{align}\label{Field Eq. Gen2}
&\left( Q_2 - \langle Q_2 \rangle \right) {\cal{K}} \nonumber\\
& = \left( Q_0 - 6 + p(p+1) + (q+1)(q-2) \right) {\cal{K}} = 0\,,
\end{align}
with the specific permissible values of the parameters $p$ and $q$ corresponding to the three series of dS UIRs.

Similar to the vector case, the above field equation only determines the \underline{physical} degrees of freedom of the elementary particles under consideration. In theories without gauge invariance, this equation perfectly captures the entire degree of freedom. However, when considering the gauge invariance properties of the dS graviton and ``partially massless'' graviton fields, the equation proves to be too restrictive. It needs to be modified to also include the gauge solutions.

For both the graviton \cite{Massless2, Massless2', Massless2'', Bamba1, dSgravity2, dSgravity1} and the ``partially massless'' graviton \cite{PMG} fields, imposing the auxiliary condition of divergencelessness introduces a singularity in the solution space. To eliminate this singularity, the condition of divergencelessness must be relaxed. Consequently, the field equation should be solved in a broader space that includes solutions where $\partial \cdot {\cal{K}} \neq 0$. This adjustment renders the field equation gauge invariant.

In the graviton (strictly massless spin-$2$) field case, associated with the discrete series UIR $(p=2 \,,\, q=2)$, or equivalently associated with the quadratic Casimir eigenvalue $\langle Q_2 \rangle = -6$, the gauge-invariant field equation explicitly reads as \cite{Massless2, Massless2', Massless2'', Bamba1, dSgravity2, dSgravity1}:
\begin{eqnarray}\label{ggg}
\left( Q_2 + 6 \right) {\cal{K}} + D_2\partial_2\cdot{\cal{K}} = 0\,,
\end{eqnarray}
with the constraint ${\cal{K}}^\prime = 0$. For any arbitrary dS vector field $\Lambda_g$:
\begin{eqnarray}
{\cal K} \mapsto {\cal K} + D_2 \Lambda_g
\end{eqnarray}
is a solution to the field equation \eqref{ggg} as long as ${\cal K}$ is. Note that, above, `$\partial_2\cdot$' refers to the generalized divergence on the dS manifold, defined as $\partial_2\cdot {\cal{K}} = \partial\cdot{\cal{K}} - R^{-2}x{\cal{K}}^\prime -\frac{1}{2}\bar\partial{\cal{K}}^\prime$,\footnote{For the traceless tensor field ${\cal{K}}$ considered here, we naturally have $\partial_2\cdot{\cal{K}} = \partial\cdot{\cal{K}}$. Additionally, for any transverse tensor field, the operator `$\partial\cdot$' acts equivalently to `$\bar\partial\cdot$'. Hence, we can extend the previous identity as follows:
\begin{align*}
    \partial_2\cdot{\cal{K}} = \partial\cdot{\cal{K}} = \bar\partial\cdot{\cal{K}}\,.
\end{align*}} and $D_2= {\cal S}(D_1 - x)$, with $D_1= R^{2}\bar\partial$. Introducing a gauge fixing parameter $c$, the graviton field then reads as:
\begin{eqnarray}\label{GravFEq}
\left( Q_2 + 6 \right) {\cal{K}} + cD_2\partial_2\cdot{\cal{K}} = 0\,.
\end{eqnarray}
Then, the solution space, rather than forming the UIR $(p=2 \,,\, q=2)$, carries an indecomposable representation of the dS group, which includes $(p=2 \,,\, q=2)$ as its central (physical) sector. For the explicit form of this indecomposable representation, refer to Ref. \cite{Massless2}.

In the ``partially massless'' graviton field case, associated with the discrete series UIR $(p=2 \,,\, q=1)$, or equivalently associated with the quadratic Casimir eigenvalue $\langle Q_2 \rangle = -4$, the gauge-invariant field equation explicitly is given by \cite{PMG}:
\begin{eqnarray}\label{pppmmm}
\left( Q_2 +4 \right) {\cal{K}} + D_2\partial_2\cdot{\cal{K}} - \theta {\cal{K}}^\prime = 0 \,, 
\end{eqnarray}
with the constraint $\partial_2\cdot {\cal{K}} = \frac{1}{2} \bar\partial{\cal{K}}^\prime$. For any arbitrary dS scalar field $\phi^{}_g$:
\begin{eqnarray}\label{Gauge}
{\cal K}\mapsto {\cal K} + D_2D_1\phi^{}_g - 2R^2 \theta \phi^{}_g
\end{eqnarray}
is a solution to the field equation \eqref{pppmmm} as long as ${\cal K}$ is. Similarly, introducing a gauge fixing parameter $c$, the field equation reads:
\begin{eqnarray}\label{Field Eq. Gen2+gauge}
\left( Q_2 +4 \right) {\cal{K}} + c D_2\partial_2\cdot{\cal{K}} - c \theta {\cal{K}}^\prime = 0\,.
\end{eqnarray}
The solution space, instead of forming the UIR $(p=2 \,,\, q=1)$, accommodates an indecomposable representation of the dS group, incorporating $(p=2 \,,\, q=1)$ as its central (physical) sector. For the explicit form of this indecomposable representation, refer to Ref. \cite{PMG}.

\emph{\textbf{Remark:} In the dS QFT literature, fields are typically described using local (intrinsic) coordinates. For a detailed explanation of the relationship between these intrinsic coordinates and the ambient ones, readers are referred to Appendix \ref{App 2}.}

\subsection{Dirac's six-cone formalism}
The so-called Dirac's six-cone is a five-dimensional surface in $\mathbb{R}^6$ \cite{Dirac1936, Mack1969, Branson1987, Binegar1983}:
\begin{align}
    (u)^2 = \eta^{}_{AB} u_{}^A u_{}^B = 0\,,
\end{align}
where $\eta^{}_{AB} = \mbox{diag}(1,-1,-1,-1,-1,1)$, with $A,B = 0,1,2,3,4,5$.

An operator $\hat{\text{O}}$, acting on scalar fields $\Phi$ in $\mathbb{R}^6$, is deemed intrinsic if, for all $\Phi$, the following holds:
\begin{align}
    \hat{\text{O}} \,(u)_{}^{2}\, \Phi = (u)^2_{}\, \hat{\text{O}}^\prime\, \Phi\,.
\end{align}
According to the given definition, the following operators are the most significant intrinsic instances:
\begin{enumerate}
    \item{The 15 conformal group SO$_0(2,4)$ generators:
    \begin{align}
        M_{AB} = -\mathrm{i} \left( u^{}_A \partial^{}_B - u^{}_B \partial^{}_A\right)\,.
    \end{align}} 
    \item{The conformal-degree operator:
    \begin{align}\label{sorati}
        \hat{N} = u^B_{} \partial^{}_B\,.
    \end{align}}
    \item{The \emph{intrinsic gradient}\footnote{Note that the ``intrinsic gradient'', introduced by Bargmann and Todorov in their seminal paper \cite{BargmannTodorov}, was originally named the ``interior derivative'' by the authors. Subsequent authors have sometimes referred to it as the ``Bargmann-Todorov operator''.}:
    \begin{align}
        \nabla_A = u^{}_A \partial^{}_B \partial_{}^B - 2 \partial^{}_A ( \hat{N} +1 ) \,.
    \end{align}}
    \item{The powers of the d'Alembertian operator, i.e., $\left(\partial^{}_B \partial^B_{}\right)^d$, act intrinsically on $\Phi$, \underline{provided} the conformal degree of $\Phi$ is $d-2$, that is, $\hat{N} \Phi = (d-2) \Phi$.}
\end{enumerate}

Given the above, we introduce the following manifestly CI system on the cone:
\begin{equation}\label{CI}
    \left\lbrace \begin{array}{lr}
        \left(\partial^{}_B \partial^B_{}\right)^d \Psi_{}^{(r)} = 0\,, \\ \\
        \hat{N} \Psi_{}^{(r)} = (d-2) \Psi_{}^{(r)} \,, 
    \end{array}\right. 
\end{equation}
where $\Psi_{}^{(r)}$ denotes a rank-$r$ tensor field with a specific symmetry (note that $\Psi^{(0)}_{} = \Phi$). To further restrict the space of solutions, particularly for \underline{symmetric} tensors, the above system can be supplemented, for instance, with the following CI conditions:
\begin{align}
    \label{Transversality} &\mbox{1. Transversality:} \quad u^{A_1} \Psi^{(r)}_{A_1 \ldots A_r} = 0\,.\\
    \label{Divergencelessness} &\mbox{2. Divergencelessness:} \quad \nabla^{A_{1}} \Psi^{(r)}_{A_1 \ldots A_r} = 0\,.\\
    \label{Tracelessness} &\mbox{3. Tracelessness:} \quad \nonumber\\
    &\quad \left(\Psi^{(r)}_{A_1\ldots A_{r}}\right)^\prime = \eta_{}^{A_{r-1}A_r} {\Psi^{(r)}_{}}^{}_{A_1\ldots A_{r-2} A_{r-1} A_r} = 0\,.
\end{align}
Note that the CI property of Eqs. \eqref{CI}, \eqref{Transversality}, \eqref{Divergencelessness}, and \eqref{Tracelessness} should be understood with respect to the invariance under the conformal infinitesimal transformations:
\begin{align}
    \delta \Psi^{(r)}_{} = \varepsilon_{}^{AB} L^{(r)}_{AB} \Psi^{(r)}_{} \,, 
\end{align}
where $L^{(r)}_{AB} = M^{}_{AB} + S^{(r)}_{AB}$, with $S^{(r)}_{AB}$ acting on the indices of $\Psi^{(r)}_{}$ in a specific permutational manner.

\subsection{Projection of the cone on dS spacetime}\label{Subsec. Projection}
The coordinates of the $(1+3)$-dimensional dS spacetime on the cone $(u)^2 = 0$ are represented by a set of five real numbers $x^a$:
\begin{align}
    x^A \equiv \left\{ x^a \;;\; (x)^2 \equiv x\cdot x = \eta^{}_{ab} x_{}^a x_{}^b = - R^2 \right\} \times \left\{ x^{}_5 \right\} \,,
\end{align}
where the following relations hold between $x^A$s and $u_{}^A$s:
\begin{align}\label{leili}
    x^a_{} = R\, \frac{u_{}^a}{u^{}_5} \,, \quad x^{}_5 = \frac{1}{R} u^{}_5\,.
\end{align}
Note that, for our purposes concerning the projective cone on dS, the fifth component $x^{}_5$ is not required. Therefore, we will continue using the notation $x$ to refer exclusively to the set $\{x^a_{}\}$.

In this context, the four intrinsic operators described earlier lead to \cite{Fronsdal1979}:
\begin{enumerate}
    \item{The conformal-degree operator:
    \begin{align}
        \hat{N}_5 = x^{}_5 \frac{\partial}{\partial x^{}_5}\,.
    \end{align}}
    \item{The (ten) dS group SO$_0(1,4)$ generators:
    \begin{align}
        M_{ab} = -\mathrm{i} \left( x^{}_a \partial^{}_b - x^{}_b \partial^{}_a\right)\,.
    \end{align}} 
    \item{The (five) purely conformal generators:
    \begin{align}\label{C generators}
        M_{5a} = -\mathrm{i} R \left( \bar\partial^{}_a + R^{-2} x^{}_a \hat{N}_5 \right)\,.
    \end{align}}
    \item{The powers of the conformal d'Alembertian operator $\left(\partial^{}_B \partial^B_{}\right)^d$, when acting on fields with conformal degree $d-2$:
    \begin{align}\label{Q(Q-2)}
        &\left(\partial^{}_B \partial^B_{}\right)^d = \nonumber\\
        &\qquad - (R\,x_5)^{-2d}\, \prod_{j=1}^d \left( Q_0 + (j+1)(j-2) \right)\,.
    \end{align}}
    \item{The conformal gradient:
    \begin{align}
        \nabla_a =&\, - \frac{1}{x^{}_5} \Big[ R^{-2} x^{}_a \left( Q_0 - \hat{N}_5\left(\hat{N}_5 -1\right) \right) \nonumber\\
        &\hspace{3.6cm} + 2\, \bar\partial^{}_a \left( \hat{N}_5 +1 \right) \Big] \,.
    \end{align}}
\end{enumerate}

As previously mentioned, a rank-$r$ tensor field $\psi^{(r)}_{a_1\ldots a_r}$ in dS spacetime is defined by its transversality and complete symmetry in its indices. The most economical way to construct such a tensor field from one residing on the cone is by projecting from a rank-$r$ symmetric tensor field $\Psi^{(r)}_{A_1\ldots A_r}$. Following the lines sketched in Ref. \cite{GazeauHans}, by contracting indices and applying a transverse projection to the latter, we introduce $(r+1)(r+2)/2$ new tensor fields as:
\begin{align}
    \Sigma^n_{n-m} \equiv \boldsymbol{\theta} \underbrace{x\cdot ( x\cdot (\ldots (x\cdot}_{m\, \text{times}} \, \Psi^{(r)}_{a_1 a_2 \ldots a_n,\, 55 \ldots 5_{(r-n\, \text{times})}}) \ldots )) \,,
\end{align}
where $0\leqslant n \leqslant r$ and $0\leqslant m \leqslant n$. The corresponding pure dS fields then are obtained by:
\begin{align}\label{sag}
    \psi^{(r)} = {x^{}_5}^{2-d} \; \Sigma^{n}_{n-m}\,.
\end{align}
Note that: (i) The role of the factor ${x^{}_5}^{2-d}$ is to eliminate the parameter $x^{}_5$ appearing in $\Sigma^{n}_{n-m}$; recall that the conformal degree of the latter, strictly speaking, $\Psi^{(r)}_{}$, is supposed to be $d-2$. (ii) The only scenario that produces a pure dS rank-$r$ tensor field occurs when $n = r$ and $m = 0$. The remaining $r(r+3)/2$ cases, i.e., $\left\{ \Sigma^n_{n-m} \,,\, n<r\; \text{or}\; n=r \;\text{and}\; m>0 \right\}$, form the set of auxiliary dS fields.

\section{CI field equations in dS spacetime}\label{Sec. III}

\subsection{Symmetric rank-$2$ tensor field $\Psi^{(2)}_{AB}$}
Let us focus specifically on the case of a symmetric rank-$2$ tensor field $\Psi^{(2)}_{AB}$ ($\Psi^{(2)}_{AB} = \Psi^{(2)}_{BA}$) defined on the cone, satisfying the set of equations \eqref{CI} with the conformal degree zero (or equivalently, $d=2$):
\begin{equation}\label{CI2}
    \left\lbrace \begin{array}{lr}
        \left(\partial^{}_B \partial^B_{}\right)^2 \Psi_{}^{(2)} = 0\,, \\ \\
        \hat{N} \Psi_{}^{(2)} = 0 \,.
    \end{array}\right. 
\end{equation}
It will soon become apparent that the choice $\hat{N} \Psi_{}^{(2)} = 0$ yields the simplest scenario that provides the unified framework we are seeking in this study.

Next, we impose the CI condition of transversality on  $\Psi_{AB}^{(2)}$. This is the only constraint we apply to the field on the cone. Without this condition, the resulting field equations become overly complex, making extracting any meaningful physical interpretation impossible. With this in place, we derive the following auxiliary CI constraints:
\begin{align} 
    \label{transverse} u_{}^A \Psi_{AB}^{(2)} = 0 \;\;&\Rightarrow\;\; x\cdot\Psi_{\cdot B}^{(2)} + R \Psi_{5B}^{(2)} = 0 \,, \\
    \label{2transverse} u_{}^A \Psi_{AB}^{(2)} u_{}^B = 0 \;\;&\Rightarrow\;\; x\cdot\Psi_{}^{(2)}\cdot x + R x\cdot \Psi_{\cdot 5}^{(2)} = 0 \,.
\end{align}
By incorporating these additional CI constraints along with \eqref{CI2} and \eqref{Q(Q-2)}, we have:
\begin{align}
    &\label{17} Q_0 \left( Q_0 - 2 \right) \Psi_{AB}^{(2)} = 0\,, \\
    &\label{28} Q_0 \left( Q_0 - 2 \right) x\cdot\Psi_{\cdot B}^{(2)} = 0\,, \\
    &\label{39} Q_0 \left( Q_0 - 2 \right) x\cdot\Psi_{}^{(2)} \cdot x = 0\,.
\end{align}

From Eqs. \eqref{17} and \eqref{28}, we respectively obtain:
\begin{align}
    &x^a_{} Q_0 \left( Q_0 - 2 \right) \Psi_{ab}^{(2)} = 0 \nonumber\\
    & \Rightarrow \;\; \left( Q_0 - 2 \right) \left( \bar\partial\cdot\Psi_{\cdot b}^{(2)} + R^{-2} x\cdot\Psi_{\cdot b}^{(2)} \right) = 0 \,, \nonumber\\
    &\label{80} \Rightarrow \;\; \bar\partial\cdot\Psi_{\cdot b}^{(2)} + R^{-2} x\cdot\Psi_{\cdot b}^{(2)} = \Lambda_b \,,
\end{align}
and:
\begin{align}
    &x^b_{} Q_0 \left( Q_0 - 2 \right) x\cdot\Psi_{\cdot b}^{(2)} = 0 \nonumber\\
    & \Rightarrow \;\; \left( Q_0 - 2 \right) \left( \bar\partial\cdot\Psi_{}^{(2)}\cdot x + R^{-2} x\cdot\Psi_{}^{(2)}\cdot x \right) = 0 \,, \nonumber\\
    &\label{90} \Rightarrow \;\; \bar\partial\cdot\Psi_{}^{(2)}\cdot x + R^{-2} x\cdot\Psi_{}^{(2)}\cdot x = \lambda\,,
\end{align}
where $\Lambda_b$ and $\lambda$ represent a vector field and scalar field, respectively, both with conformal degree zero and satisfy $\left( Q_0 - 2 \right) \Lambda_b = 0$ and $\left( Q_0 - 2 \right) \lambda = 0$. For simplicity, we further impose on $\Lambda_b$ the transversality condition ($x \cdot \Lambda = 0$) and the divergenceless condition ($\bar\partial \cdot \Lambda = 0$). These constraints allow us to associate $\Lambda_b$ with the discrete UIR $(p = 1 \,,\, q = 1)$. To aid future analysis, we now examine the consequences of Eqs. \eqref{80} and \eqref{90} in more detail:
\begin{align}
    &x^b_{}\left( \bar\partial\cdot\Psi_{\cdot b}^{(2)} + R^{-2} x\cdot\Psi_{\cdot b}^{(2)} = \Lambda_b \right) \; \text{and Eq. \eqref{90}} \nonumber\\
    \label{100} & \Rightarrow \;\; \eta^{ab}_{}\Psi_{ab}^{(2)} + R^{-2} x\cdot\Psi_{}^{(2)}\cdot x = \lambda\,.
\end{align}

\subsection{Reduction to dS spacetime}
In this context, following the steps sketched in Sect. \ref{Subsec. Projection}, the following six dS fields come to the fore:
\begin{align}
    \Sigma^{2}_{2} =&\, \boldsymbol{\theta}\, \Psi_{a b}^{(2)} = \theta_a^{a^\prime} \theta_b^{b^\prime} \Psi_{a^\prime b^\prime}^{(2)}\,,\\ 
    \Sigma^{2}_{1} =&\, \boldsymbol{\theta}\, x\cdot \Psi_{\cdot a}^{(2)} = \theta_a^{a^\prime} x\cdot \Psi_{\cdot a^\prime}^{(2)} \,, \\ 
    \Sigma^{2}_{0} =&\, x\cdot \Psi_{}^{(2)} \cdot x \,, \\
    \Sigma^{1}_{1} =&\, \boldsymbol{\theta}\, \Psi_{a 5}^{(2)} = \theta_a^{a^\prime} \Psi_{a^\prime 5}^{(2)}\,, \\
    \Sigma^{1}_{0} =&\, x\cdot \Psi_{\cdot 5}^{(2)}\,, \\ 
    \Sigma^{0}_{0} =&\, \Psi_{55}^{(2)}\,. 
\end{align}
Interestingly, imposing the CI constraints \eqref{transverse}, \eqref{2transverse}, \eqref{80}, \eqref{90}, and \eqref{100} allows us to establish the following relations:
\begin{align}
    \Sigma^2_1 =&\, - R\Sigma^1_1\,, \\
    \Sigma^2_0 =&\, - R\Sigma^1_0\,, \\
    \label{turace} \left(\Sigma^2_2\right)^\prime =&\, \lambda \,,\\
    \label{jir00} \bar\partial \cdot \Sigma^2_2 =&\, 4R^{-2} \Sigma^2_1 + \Lambda + R^{-2} x\lambda \,.
\end{align}
Upon plugging $\lambda$ from Eq. \eqref{turace} into Eq. \eqref{jir00}, we obtain:
\begin{align}
    \label{jir} \bar\partial \cdot \Sigma^2_2 =&\, 4R^{-2} \Sigma^2_1 + \Lambda + xR^{-2} \left(\Sigma^2_2\right)^\prime \,,
\end{align}
which subsequently leads to:
\begin{align}
    \bar\partial \cdot \bar\partial \cdot \Sigma^2_2 =&\, 4R^{-2} \left(\bar\partial \cdot \Sigma^2_1 \right) + 4R^{-2} \left(\Sigma^2_2\right)^\prime \nonumber\\
    \label{wqwq} =&\, 4R^{-2} \left(3R^{-2} \Sigma^2_0 + \lambda \right) + 4R^{-2} \left(\Sigma^2_2\right)^\prime \,.
\end{align}

Given the above relations, we now focus exclusively on the following cases:
\begin{enumerate}
    \item{The rank-$2$ tensor field ${\cal{K}} \equiv \Sigma^2_2$,}
    \item{The vector field $K \equiv \Sigma^2_1$.}
\end{enumerate} 
We examine these dS fields in detail in the following subsections. For the sake of completeness, however, it is useful to briefly mention the CI equations governing the remaining fields:
\begin{enumerate}
    \item{Based on Eqs. \eqref{transverse} and \eqref{2transverse}, the vector field $\Sigma^1_1$ exhibits behavior closely analogous to that of the vector field \( K \equiv \Sigma^2_1 \). In particular, it satisfies the same CI equation as \( K \), which will be introduced in the following discussion.}
    \item{On the other hand, based on Eqs. \eqref{17}–\eqref{39}, the remaining scalar fields, namely \( \Sigma^2_0 \), \( \Sigma^1_0 \), and \( \Sigma^0_0 \), are governed by the following common CI field equation:
    \begin{align}
        Q_0 \left( Q_0 - 2 \right) \Sigma^2_0 =&\, 0 \,, \\
        Q_0 \left( Q_0 - 2 \right) \Sigma^1_0 =&\, 0 \,, \\
        Q_0 \left( Q_0 - 2 \right) \Sigma^0_0 =&\, 0 \,.
    \end{align}}
\end{enumerate}


\subsection{Vector field $K \equiv \Sigma^2_1$}
We begin by examining the dS vector field:
\begin{align}
    K \equiv \Sigma^2_1 = x\cdot \Psi^{(2)}_{\cdot a} + R^{-2} x^{}_a x\cdot \Psi^{(2)}_{}\cdot x \,, \label{159}
\end{align}
while:
\begin{align}
    \bar\partial\cdot K = 3R^{-2} x\cdot \Psi^{(2)}_{}\cdot x + \lambda \,. \label{951}
\end{align}
Utilizing these definitions, Eqs. \eqref{28} and \eqref{39}, and the fact that $\left( Q_0 - 2 \right)\lambda = 0$, we derive:
\begin{align}\label{sok}
    &Q_0 \left( Q_0 - 2 \right) K_a = \frac{1}{3} Q_0 \left( Q_0 - 2 \right) x^{}_a \left(\bar\partial\cdot K - \lambda \right) \nonumber\\
    & \hspace{2cm} = -\frac{4}{3} \left(3x^{}_a + D_{1a} \right) \left( Q_0 - 2 \right) \bar\partial\cdot K \,,
\end{align}
with the constraint:
\begin{align}\label{sok'}
    Q_0 \left( Q_0 - 2 \right) \bar\partial\cdot K = 0 \,.
\end{align}
Note that the above equations are invariant under the purely conformal transformation $K \mapsto K + \delta K$, where (see Eq. \eqref{C generators}; noting that $\hat{N}_5 K = 0$ as $\hat{N}_5 \Psi^{(2)}_{} = 0$):
\begin{align}\label{sadafi}
    \delta K =&\, - \mathrm{i} R \left( \boldsymbol{\theta} Z\cdot\bar\partial K \right) \nonumber\\
    =&\, - \mathrm{i} R \left( Z\cdot\bar\partial K - R^{-2} x(Z\cdot K) \right)\,,
\end{align}
with $Z_a$ bieng the infinitesimal five-vector $Z_a = \varepsilon^{}_{5a}$.

Through a series of straightforward computations (detailed in Appendix \ref{App A}), it can be shown that Eq. \eqref{sok}, subject to the condition \eqref{sok'}, simplifies to:
\begin{align}\label{sok 5}
    Q_1 K_a + \frac{1}{6} D_{1a} \left( Q_0 + 4 \right) \bar\partial\cdot K = 0 \,,
\end{align}
which retains its invariance under the purely conformal transformation described in Eq. \eqref{sadafi}. Notably, this equation parallels Eq. \eqref{masslessvector}, derived from a purely dS group-theoretical framework, as we will explore further in Sect. \ref{Sec. conclusion}.

\subsection{Rank-$2$ tensor field ${\cal{K}} \equiv \Sigma^2_2$}
By definition, the dS rank-$2$ tensor field is given by: 
\begin{align}
    {\cal{K}} \equiv \Sigma^2_2 = \Psi_{a b}^{(2)} + R^{-2} {\cal{S}} x^{}_a x\cdot \Psi^{(2)}_{\cdot b} + R^{-4} x^{}_a x^{}_b x\cdot \Psi^{(2)}_{}\cdot x\,. 
\end{align}
We decompose \underline{in a CI manner} the tensor field $\mathcal{K}_{ab}$ as the sum of a traceless part and a pure-trace part:
\begin{align}\label{cedompo}
    \mathcal{K}^{}_{ab} = \mathcal{K}^0_{ab} + \frac{1}{4} \theta^{}_{ab} \mathcal{K}^\prime \,.
\end{align}
By applying Eqs. \eqref{17}–\eqref{39}, along with \eqref{jir} and \eqref{wqwq}, and performing some straightforward computations, we arrive at:
\begin{widetext}
\begin{align}
    Q_0 \left( Q_0 - 2 \right) {\cal{K}}^0_{ab} =&\, {\frac{1}{24}} Q_0 \left( Q_0 - 2 \right) {\cal{S}} \left[ 6 x^{}_a \bar\partial\cdot{\cal{K}}^0_{\cdot b} - x^{}_a x^{}_b \bar\partial\cdot\bar\partial\cdot{\cal{K}}^0_{} \right]\nonumber \\
    =&\, \frac{1}{6} {\cal{S}} \big[ -2 D_{1a} (3x^{}_b + D_{1b}) \left( Q_0 - 2 \right) + 3 (3x^{}_a + D_{1a}) \left( Q_0 - 2 \right) (3x^{}_b + D_{1b}) \nonumber\\
    &\quad\quad - 7 x^{}_a(3x^{}_b + D_{1b}) \left( Q_0 - 2 \right) - 8 (3x^{}_a + D_{1a}) \left( Q_0 - 2 \right) x^{}_b \big] \bar\partial\cdot\bar\partial\cdot{\cal{K}}^0_{}
    \label{tt}\,,\\
    Q_0 \left( Q_0 - 2 \right) \theta_{ab} {\cal{K}}^\prime_{} =&\, -2 R^{-2} {\cal{S}} \left[ x^{}_a (3x^{}_b + D_{1b} ) \left( Q_0 - 2 \right) + ( 3x^{}_a + D_{1a} ) \left( Q_0 - 2 \right) x^{}_b \right] {\cal{K}}^\prime\,, \label{pt} 
\end{align}
with the constraints:
\begin{align}
    \label{56} Q_0 \left( Q_0 - 2 \right) \bar\partial\cdot\bar\partial\cdot{\cal{K}} =&\; 0 = Q_0 \left( Q_0 - 2 \right) \bar\partial\cdot\bar\partial\cdot{\cal{K}}^0_{}\,, \\
    \label{65} Q_0 \left( Q_0 - 2 \right) {\cal{K}}^\prime =&\; 0\,.
\end{align}
Note that: (i) To obtain Eq. \eqref{tt}, we have also used the fact that $Q_0 \left( Q_0 - 2 \right)x\Lambda = 0$ and ${\cal{K}}^\prime = \lambda$. (ii) In the particular traceless case (where $\bar{\partial} \cdot \mathcal{K}^0_{\cdot a} = 4R^{-2} K^{}_a$; the term $\Lambda_a$ is assumed to couple with $K^{}_a$), multiplying Eq. \eqref{tt} by $x^a_{}$ yields a result that precisely matches Eq. \eqref{sok}, and subsequently, \eqref{sok 5}. (iii) Similarly, the consistency of Eq. \eqref{pt} can be easily confirmed by multiplying both sides by $\eta^{ab}_{}$, which directly reduces to Eq. \eqref{65}. (iv) The above equation remains invariant under the purely conformal transformation ${\cal{K}} \mapsto {\cal{K}} + \delta {\cal{K}}$, where (see Eq. \eqref{C generators}; noting that $\hat{N}_5 {\cal{K}} = 0$ since $\hat{N}_5 \Psi^{(2)}_{} = 0$):
\begin{align} \label{diksy}
    \delta {\cal{K}} = - \mathrm{i} R \left( \boldsymbol{\theta} Z\cdot\bar\partial {\cal{K}} \right) = - \mathrm{i} R \left( Z\cdot\bar\partial {\cal{K}} - R^{-2} {\cal{S}} x(Z\cdot {\cal{K}}) \right)\,,
\end{align}
where, as before, $Z_a$ represents the infinitesimal five-vector $Z_a = \varepsilon^{}_{5a}$. (v) Regarding the latter point, it is worth noting that the dS traceless condition, ${\cal{K}}^\prime = 0$, remains invariant under the purely conformal transformation given by Eq. \eqref{diksy}, which ensures that the decomposition in Eq. \eqref{cedompo} is CI. (vi) Combining Eqs. \eqref{tt} and \eqref{pt} yields the CI equation for the full traceful field $\mathcal{K}$:
\begin{align}
    Q_0 \left( Q_0 - 2 \right) {\cal{K}}^{}_{ab} =&\, \frac{1}{6} {\cal{S}} \big[ -2 D_{1a} (3x^{}_b + D_{1b}) \left( Q_0 - 2 \right) + 3 (3x^{}_a + D_{1a}) \left( Q_0 - 2 \right) (3x^{}_b + D_{1b}) \nonumber\\
    &\quad\quad - 7 x^{}_a(3x^{}_b + D_{1b}) \left( Q_0 - 2 \right) - 8 (3x^{}_a + D_{1a}) \left( Q_0 - 2 \right) x^{}_b \big] \left( \bar\partial\cdot\bar\partial\cdot{\cal{K}} - 4R^{-2}{\cal{K}}^\prime + \frac{1}{4} R^{-2} Q_0 {\cal{K}}^\prime \right) \nonumber\\
    &\quad\quad - \frac{1}{2} R^{-2} {\cal{S}} \left[ x^{}_a (3x^{}_b + D_{1b} ) \left( Q_0 - 2 \right) + ( 3x^{}_a + D_{1a} ) \left( Q_0 - 2 \right) x^{}_b \right] {\cal{K}}^\prime\,.
\end{align}

By performing a series of straightforward calculations (outlined in Appendix \ref{App B}), it can be demonstrated that Eqs. \eqref{tt} and \eqref{pt}, subject to the constraints \eqref{56} and \eqref{65}, simplify to:
\begin{align}\label{pmg-ci}
    &\left( Q_2 + 4 \right) {\cal{K}}^{}_{ab} + \frac{2}{3} D_{2a} \partial_2\cdot{\cal{K}}^{}_{\cdot b} + \frac{1}{6}D_{2a} \bar\partial_b {\cal{K}}^\prime - \frac{1}{3} R^2 \theta_{ab} \bar\partial\cdot\partial_2\cdot{\cal{K}} - \frac{1}{6} \theta_{ab} \left(Q_0 + 6 \right) {\cal{K}}^\prime \nonumber\\
    &\hspace{9cm} = \frac{1}{108} D_{2a}D_{1b}\left( Q_0 - 2 \right) \left(\bar\partial\cdot\bar\partial\cdot{\cal{K}} - 16R^{-2}{\cal{K}}^\prime \right)\,,
\end{align}
and:
\begin{align}\label{ttGsolu2}
    \left( Q_2 + 6 \right) {\cal{K}}^{0}_{ab} + D_{2a} \bar\partial\cdot{\cal{K}}^0_{\cdot b} = \frac{1}{72} \left( 32 R^2 \theta_{ab} - 4D_{2a} D_{1b} \right)\bar\partial\cdot\bar\partial\cdot{\cal{K}}^0_{} + \frac{1}{72} \left( D_{2a} D_{1b} - 2R^2\theta_{ab} \right) \left( Q_0 - 2 \right) \bar\partial\cdot\bar\partial\cdot{\cal{K}}^0_{} \,.
\end{align}
\end{widetext}
Note that these equations should be, respectively, compared with Eqs. \eqref{Field Eq. Gen2+gauge} and \eqref{GravFEq}, which were obtained through a purely dS group-theoretical mechanism, as we will elaborate further in Sect. \ref{Sec. conclusion}.

\textit{\textbf{Remark:} One may observe that equations \eqref{pmg-ci} and \eqref{ttGsolu2} exhibit an intriguing feature: they do not possess a smooth flat-spacetime limit. Specifically, when the dependence on the dS radius $R$ is made explicit, it becomes clear that these equations do not admit a regular transition to flat Minkowski spacetime. This behavior, however, is not unexpected, as these equations are associated with the descriptions of the ``partially massless'' and strictly massless graviton fields, respectively (see the following section).}

\textit{It is worth recalling that the dS group admits no UIRs analogous to the so-called massless infinite-spin representations of the Poincar\'{e} group. In this context, massless representations of the dS group are naturally identified as those that admit a unique extension to UIRs of the conformal group SO$_0(2,4)$, matching the conformal extensions of the massless UIRs of the Poincar\'{e} group.}

\textit{It should also be noted that the ``partially massless'' graviton field has no Minkowskian counterpart, not even through the conformal extension described above. Therefore, the absence of a meaningful flat-spacetime limit should not come as a surprise.}

\textit{\textbf{Remark:} It is important to emphasize that, unlike the earlier fourth-order differential equations \eqref{tt} and \eqref{pt}, which preserve conformal invariance, the second-order differential equations \eqref{pmg-ci} and \eqref{ttGsolu2} are not conformally invariant.}

\textit{\textbf{Remark:} To conclude this section, it is also important to note that Eqs. \eqref{sok 5} and \eqref{tt} are comparable to the results presented in Ref. \cite{Takook}, which were obtained through a different technical approach. This alternative method involves a more restrictive framework, imposing both divergenceless and traceless conditions. In this regard, it is worth noting that the use of these constraints on the conformal graviton field in Ref. \cite{Takook} prevents the CI equations in dS spacetime from being uplifted to a manifestly six-dimensional CI form.}

\section{Summary and discussion}\label{Sec. conclusion}
Utilizing Dirac's six-cone formalism, we began with a symmetric, transverse, rank-$2$ tensor field $\Psi^{(2)}_{AB}$ ($A, B = 0,\ldots,5$) of conformal degree zero on the cone. We explored its reduction to the dS manifold. Among the six possible dS-reductive fields, we specifically focused on the rank-$1$ (vector) field $K_a \equiv \boldsymbol{\theta} \, x \cdot \Psi^{(2)}_{\cdot a}$ and the rank-$2$ tensor field ${\cal{K}}_{ab} \equiv \boldsymbol{\theta} \Psi^{(2)}_{ab}$ ($a, b = 0, \dots, 4$). This led to a set of \underline{compatible} CI fourth-order differential equations: 
\begin{enumerate}
    \item{Eq. \eqref{sok} for $K$, subject to the CI constraint \eqref{sok'},}
    \item{Eqs. \eqref{tt} and \eqref{pt} for ${\cal{K}}$, subject to the CI constraints \eqref{56} and \eqref{65}.}
\end{enumerate}

We argue that this CI framework provides a unified setup encompassing three elementary fields in dS spacetime: the massless vector (photon) field (see Sect. \ref{Subsect. 1}), the ``partially massless'' graviton field (see Sect. \ref{Subsect. 2}), and the graviton (strictly massless spin-$2$) field (see Sect. \ref{Subsect. 2}). Specifically, under certain \underline{consistent} conditions \underline{involving conformal symmetry breaking}, this CI system gives rise to the aforementioned fields, as we further explain now.

\subsection{The massless vector (photon) field}
We demonstrated that the CI fourth-order differential equation \eqref{sok} for $K$ in CI way reduces to the second-order differential equation \eqref{sok 5}. A comparison with its counterpart, Eq. \eqref{masslessvector}, which is derived using a purely dS group-theoretical approach for the massless vector (photon) field in dS spacetime, reveals that the two equations coincide if either of the following \underline{consistent} conditions is met:\footnote{Again, for a transverse dS tensor field, such as $K$, we have $\bar\partial\cdot K = \partial\cdot K$.}
\begin{align}
    \label{001} Q_0 \bar\partial \cdot K =&\, 0\,,\\
    \label{002} \left( Q_0 - 2 \right) \bar\partial \cdot K =&\, 0\,.
\end{align}
Note that these equations are indeed \underline{consistent} with the CI constraint \eqref{sok'}, though they are not themselves CI (in the sense of \eqref{sadafi}). Imposing these non-CI conditions on Eq. \eqref{sok 5} yields the following results, respectively:
\begin{align}
    \label{001'} Q_1 K_a + \frac{2}{3} D_{1a} \bar\partial\cdot K = 0 \,,\\
    \label{002'} Q_1 K_a + D_{1a} \bar\partial\cdot K = 0 \,.
\end{align}
These conditions correspond to the gauge fixings $c = 2/3$ and $c = 1$, respectively, in the twin equation \eqref{masslessvector}. The first case, known as the minimal choice for the gauge-fixing parameter, eliminates logarithmic divergences in the space of field solutions, which would otherwise cause reverberations inside the light cone. The scenario $c \neq 1$ has been thoroughly studied by Gazeau et al. in Ref. \cite{Gazeaus1}. Although the second case ($c = 1$) may initially appear fully gauge-invariant — and thus unsuitable for constructing a covariant QFT formulation of the corresponding field — it is, in fact, constrained by the condition \eqref{002}, which limits the gauge-field space. This case has not yet been explored in the literature and merits further investigation, especially given that both Eqs. \eqref{001'} and \eqref{002'} arise from conformal symmetry considerations. 

\subsection{The ``partially massless'' graviton field}
We also showed that the CI fourth-order differential equations \eqref{tt} and \eqref{pt} for ${\cal{K}}$ simplify to two second-order (non-CI) differential equations, \eqref{pmg-ci} and \eqref{ttGsolu2}. In this subsection, we focus on the former case, with a detailed discussion of the latter deferred to the following subsection.

A direct comparison of Eq. \eqref{pmg-ci} with its twin, Eq. \eqref{Field Eq. Gen2+gauge}, which is derived using a purely dS group-theoretical approach for the ``partially massless'' graviton in dS spacetime, shows that the two equations coincide if the following conditions are imposed:
\begin{align}
    \label{003} &\partial_2 \cdot \mathcal{K} = \frac{1}{2} \bar{\partial} \mathcal{K}^\prime\,, \; (\mbox{equiv.,} \; \bar{\partial} \cdot \mathcal{K} = \bar{\partial} \mathcal{K}^\prime + R^{-2} x \mathcal{K}^\prime)\,, \\
    \label{004} &\left( Q_0 - 2 \right) \bar\partial\cdot\bar\partial\cdot {\cal{K}} = 0 \,,\\
    \label{005} & \left( Q_0 - 2 \right) {\cal{K}}^\prime = 0 \,.
\end{align}
Note that the first constraint is consistent with Eq. \eqref{jir}, as it partially restricts the degrees of freedom associated with $\Lambda$. The remaining constraints are in agreement with the CI conditions \eqref{56} and \eqref{65}. Moreover, one should notice that the constraint \eqref{003} precisely coincides with the very condition previously introduced to derive Eq. \eqref{Field Eq. Gen2+gauge}. By imposing these constraints, Eq. \eqref{pmg-ci} takes the following form:
\begin{align}\label{tt+pt}
    \left( Q_2 + 4 \right) {\cal{K}}^{}_{ab} + D_{2a} \partial_2\cdot{\cal{K}}^{}_{\cdot b} - \theta_{ab} {\cal{K}}^\prime = 0\,.
\end{align}
The latter corresponds to the gauge fixing $c=1$ in the twin Eq. \eqref{Field Eq. Gen2+gauge}. It is important to emphasize that this scenario does \underline{not} represent a fully gauge-invariant case, as the gauge-field space is restricted not only by condition \eqref{003}, which already exists in the dS description of the field, but also by the additional conditions \eqref{004} and \eqref{005}. While we previously studied the field equation for the gauge fixing $c\neq 1$ (subject only to \eqref{003}) in Ref. \cite{PMG}, the case $c = 1$ (with the additional conditions \eqref{004} and \eqref{005}), which is derived from the conformal symmetry properties of the field, remains unexplored and merits a thorough examination.

\emph{\textbf{Remark:} In a forthcoming work, we aim to conduct a comprehensive investigation of the phase-space realization of the ``partially massless'' graviton field, followed by an investigation of its covariant integral quantization, as has been done, for instance, in the $1+1$-dimensional anti-dS case \cite{GazOlmo}. Furthermore, studying the backreaction of this field, together with the ``massless'' minimally coupled scalar field, on the metric within a semiclassical framework involving the dS spacetime promises to be particularly intriguing.}

\subsection{The graviton (strictly massless spin-$2$) field}
As mentioned in the previous subsection, another reduction of the CI fourth-order differential equation \eqref{tt} is achieved through the second-order (non-CI) differential equation \eqref{ttGsolu2}. A comparison with its counterpart, Eq. \eqref{GravFEq},\footnote{Note that, for clarity of reasoning, the traceless condition on the tensor field in Eq. \eqref{GravFEq} is treated as an additional constraint, ${\cal{K}}^\prime = 0$, rather than being incorporated directly into the equation by adding a superscript `$^0$' to the field.} derived using a purely dS group-theoretical approach for the strictly massless graviton field in dS spacetime, shows that the two equations coincide when the following consistent condition is met:\footnote{Again, for a transverse and traceless rank-$2$ tensor field ${\cal{K}}$, we have $\partial_2\cdot {\cal{K}} = \partial\cdot{\cal{K}} = \bar\partial\cdot{\cal{K}}$.}
\begin{align}
    \label{006} \bar\partial\cdot\bar\partial\cdot {\cal{K}}^0_{} =&\, 0 \,.
\end{align}
Although this constraint is not CI, it is indeed consistent with the CI constraint \eqref{56}. Taking these constraints into account, Eq. \eqref{ttGsolu2} simplifies to:
\begin{align}\label{5849678}
    \left( Q_2 + 6 \right) {\cal{K}}^{0}_{ab} + D_{2a} \partial_2\cdot{\cal{K}}^{0}_{\cdot b} = 0\,.
\end{align}
This equation corresponds to the gauge fixing $c = 1$ in the twin equation \eqref{GravFEq}. However, it is important to note that this scenario does \underline{not} represent a fully gauge-invariant case, as the gauge-field space is constrained by the condition \eqref{006}. While we previously examined the field equations for gauge fixing with $c \neq 1$ in Refs. \cite{Massless2, Massless2', Massless2'', Bamba1, dSgravity2, dSgravity1}, the case $c = 1$ (equipped with \eqref{006}), which arises from the conformal symmetry properties of the field, remains unexplored and warrants a thorough investigation.

\emph{\textbf{Remark:} Although the above construction does not explicitly address the massless scalar field governed by the field equation $(Q_0 - 2)\phi = 0$ (see Sect. \ref{Sect. scalar}), the operator $(Q_0 - 2)$ plays a critical role that can be traced throughout the framework.}

\subsection{The importance of global symmetry in dS physics}  
As a concluding point, we highlight an important aspect of dS physics that may be underappreciated in observer-dependent approaches: the unique insights afforded by the global symmetry framework. While the static patch approach, commonly adopted in cosmology, is practical for modeling local observables within the constraints of a cosmological event horizon, it inherently sacrifices the full SO$_0(1,4)$ symmetry of the dS manifold. This reduction to the SO$_0(1,3)$ symmetry of the static patch limits access to the unifying perspective provided by global symmetry.

The global framework, as emphasized in this work, is far from a purely formal or abstract construct. It encapsulates critical aspects of dS physics, including the thermodynamic properties of the horizon, the dynamics of quantum fields, and the compatibility of quantum gravity with conformal symmetry. Specifically, the discussion in this paper centers on Dirac's six-cone formalism, which provides a unified treatment of massless and ``partially massless'' fields in dS spacetime under conformal invariance. This formalism inherently depends on the global structure and cannot be meaningfully reduced to specific patches. This underscores the indispensable role of global symmetry in preserving the coherence of the physical framework.

Moreover, recent research by the authors \cite{EPL} highlights the crucial role of global symmetry in maintaining the internal consistency of dS spacetime. This study reveals that the matter-antimatter asymmetry, often regarded as a fundamental cosmological challenge, can manifest as an observer-dependent effect arising from the time orientation within local causal patches. This asymmetry does not stem from an intrinsic property of dS spacetime but emerges due to the limitations of local perspectives. By leveraging the unique causal and analytic structure of the dS manifold, the study demonstrates that global symmetry considerations reconcile this apparent asymmetry without contradicting established mechanisms such as baryon number violation and CP violation. This exemplifies how certain physical effects may appear differently—or even paradoxically—when analyzed solely within a local patch.

It is, however, essential to clarify that this emphasis on global symmetry does not diminish the value of the static patch approach. On the contrary, the two perspectives should be regarded as complementary. The global viewpoint provides a comprehensive framework for understanding symmetries and field dynamics on the entire manifold, while the static patch remains indispensable for exploring locally accessible phenomena, particularly those relevant to cosmological observations. Together, they form a multifaceted toolkit that enables a richer and more complete understanding of dS physics from different, yet compatible, vantage points.

\section*{Acknowledgements}
Hamed Pejhan extends his gratitude to Ivan Todorov for his insightful comments and support throughout the completion of this project. Hamed Pejhan is supported by the Bulgarian Ministry of Education and Science, Scientific Programme ``Enhancing the Research Capacity in Mathematical Sciences (PIKOM)'', No. DO1-67/05.05.2022. Jean-Pierre Gazeau expresses his gratitude to the Institute of Mathematics and Informatics, Bulgarian Academy of Sciences, for hosting his visit under the PIKOM programme, during which the finalization of this work was accomplished.

\setcounter{equation}{0} 
\begin{appendix}
\section{Some useful relations}
Here are some important relations crucial for verifying the equations presented in this article:
\begin{eqnarray}\label{Identities}
\partial_2\cdot\theta\phi &=& -R^{-2}D_1 \phi\,,\\
Q_1 D_1 \phi &=& D_1 Q_0 \phi\,, \\
\partial_2\cdot D_2D_1 \phi &=& -\left(Q_1 +6 \right) D_1 \phi \nonumber\\
&=& - D_1 \left(Q_0 +6\right) \phi\,,
\end{eqnarray}
where $\phi$ is a scalar field on dS spacetime.

Moreover, we have the following relations:
\begin{align}
    \label{xxxx} \left( Q_0 - 2 \right) x^{}_a =&\, x^{}_a \left( Q_0 - 2 \right) - 4x^{}_a - 2 D_{1a}\,,\\
    \label{dddd} \left( Q_0 - 2 \right) D_{1a} =&\, D_{1a} \left( Q_0 - 2 \right) + 6D_{1a} \nonumber\\
    &\, + 2 \left(Q_0 + 4\right) x^{}_a\,,\\
    Q_0\left( Q_0 - 2 \right) x^{}_a =&\, x^{}_a Q_0 \left( Q_0 - 2 \right) \nonumber\\
    &\, - 4 (3x^{}_a + D_{1a}) \left( Q_0 - 2 \right) \,,\\
    Q_0\left( Q_0 - 2 \right) D_{1a} =&\, (4x^{}_a + D_{1a}) \; Q_0 \left( Q_0 - 2 \right) \,, \\
    D_{1a} D_{1b} =&\, D_{1b} D_{1a} + x^{}_b D_{1a} - x^{}_a D_{1b}\,.
\end{align}

\section{Link to intrinsic coordinates}\label{App 2}
In dS QFT literature, fields are commonly described using local (intrinsic) coordinates. Therefore, it is valuable to establish the connection between these intrinsic coordinates and the ambient ones. For clarity, we will focus exclusively on the relations relevant to our study, particularly those concerning a rank-$2$ tensor field and its field equation.

The following identity defines the relation between ${\cal{K}}_{ab}(x)$ and its (local) intrinsic counterpart $h_{\mu\nu}(X)$:
\begin{eqnarray}\label{samir}
h_{\mu\nu}(X) = x^{a}_{\,\,,\,\mu} x^{b}_{\,\,,\,\nu} \; {\cal{K}}_{ab} \big(x(X)\big)\,,
\end{eqnarray}
where $x^{a}_{\,\,,\,\mu} = \partial x^{a}/\partial X^{\mu}$, while $X^\mu$ ($\mu=0,1,2,3$) denotes the four local spacetime coordinates on dS. Inducing the natural ambient Minkowski ($\mathbb{R}^5$) metric onto the dS manifold yields the metric on the dS manifold:
\begin{eqnarray}
\mathrm{d} s^2= \eta_{ab}\mathrm{d} x^{a}\mathrm{d} x^{b}\big|_{(x)^2=-R^{2}} = g_{\mu\nu} \mathrm{d} X^{\mu}\mathrm{d} X^{\nu}\,.
\end{eqnarray}
Considering the identity \eqref{samir}, $\theta^{}_{ab}$ is the unique symmetric and transverse tensor associated with the dS metric, where $g_{\mu\nu}= x^a_{\,\,,\,\mu} x^b_{\,\,,\,\nu} \theta^{}_{ab}$. 

The following transformation applies to the covariant derivatives:
\begin{eqnarray}
\nabla_\rho \nabla_\lambda h_{\mu\nu} = x_{\,\,,\,\rho}^{c}x_{\,\,,\,\lambda}^{d} \;\; x_{\,\,,\,\mu}^{a}x_{\,\,,\,\nu}^{b} \;\; ({\boldsymbol{\theta}}\bar\partial_c) ({\boldsymbol{\theta}}\bar\partial_d) \; {\cal{K}}_{ab}\,,
\end{eqnarray}
where, for a given symmetric rank-$r$ tensor field ${\psi}^{(r)}_{a_1 \,...\, a_r}(x)$, the operator ${\boldsymbol{\theta}}$ is defined as:
\begin{align}
    \left( \prod_{i=1}^r \theta^{b_i}_{a_i} \right) {\psi}^{(r)}_{b_1 \,...\, b_r}(x) \equiv ({\boldsymbol{\theta}} {\psi})^{(r)}_{a_1 \,...\, a_r}(x)\,.
\end{align}
By construction, ${\boldsymbol{\theta}}$ ensures the transversality of the field in each tensorial index.

In this context, the d'Alembertian operator is linked to the scalar Casimir operator $Q_0$ (see also Eq. \eqref{Q_0}):
\begin{eqnarray}
\square_R \phi = g^{\mu\nu} \nabla_\mu \nabla_\nu \phi &=& g^{\mu\nu} x^a_{\,\,,\,\mu} x^b_{\,\,,\,\nu} \Big( \bar\partial_a \bar\partial_b - R^{-2} x^{}_b \bar\partial_a \Big)\phi \nonumber\\
&=& \theta^{ab} \Big( \bar\partial_a \bar\partial_b - R^{-2} x^{}_b \bar\partial_a \Big)\phi \nonumber\\
&=& \bar\partial^2 \; \phi = -R^{-2} Q_0\,,
\end{eqnarray}
where $\phi$ denotes a dS scalar field.

\section{Derivation and detailed calculations for Eq. \eqref{sok 5}} \label{App A}
For the sake of reasoning, let $\mathfrak{S}$ be a three-dimensional space spanned by linear combinations of the following set of three functions:
\begin{align}
    \mathfrak{S} \ni [ \boldsymbol{s_1}, \boldsymbol{s_2}, \boldsymbol{s_2} ] =\, &\boldsymbol{s_1} \left[\left(3x^{}_a + D_{1a} \right) \left( Q_0 - 2 \right) \bar\partial\cdot K\right] \nonumber\\ 
    +\, & \boldsymbol{s_2} \left[\left( Q_0 - 2 \right) x^{}_a \bar\partial\cdot K\right] \nonumber\\ 
    +\, & \boldsymbol{s_3} \left[\left( Q_0 - 2 \right) D_{1a} \bar\partial\cdot K\right] \,.
\end{align}
Notably, this space remains invariant under the action of $Q_0$:
\begin{align}
    Q_0 \left[\left(3x^{}_a + D_{1a} \right) \left( Q_0 - 2 \right) \bar\partial\cdot K\right] =&\, [ -4, 0, 0 ]\,,\\
    Q_0 \left[\left( Q_0 - 2 \right) x^{}_a \bar\partial\cdot K\right] =&\, [ -4, 0, 0 ]\,, \\
    Q_0 \left[\left( Q_0 - 2 \right) D_{1a} \bar\partial\cdot K\right] =&\, [ 0, 0, 0 ]\,.
\end{align}

With respect to this three-dimensional space, we can rewrite Eq. \eqref{sok} as:
\begin{align}\label{sok 1}
    Q_0 \left( Q_0 - 2 \right) K_a = \left[-\frac{4}{3}, 0 , 0 \right] \,.
\end{align}
This equation can then be simplified to: \begin{align}\label{sok 2}
    \left( Q_0 - 2 \right) K_a = [ \boldsymbol{s_1}, \boldsymbol{s_2}, \boldsymbol{s_3} ] \,,
\end{align}
provided that:
\begin{eqnarray}
\left(\begin{array}{cccccccc}
-4 & -4 & 0 \\
0 & 0 & 0 \\
0 & 0 & 0 
\end{array}\right)
\left(\begin{array}{cccccccc}
\boldsymbol{s_1} \\
\boldsymbol{s_2} \\
\boldsymbol{s_3} 
\end{array}\right) =
\left(\begin{array}{cccccccc}
-{4}/{3} \\
0 \\
0
\end{array}\right)\,,
\end{eqnarray}
which implies that:
\begin{align}\label{sok 3}
    \boldsymbol{s_1} + \boldsymbol{s_2} = \frac{1}{3} \,.
\end{align}
On the other hand, multiplying both sides of Eq. \eqref{sok 2} by $x^a_{}$ and/or $\bar\partial^a_{}$ readily confirms that the consistent solution must satisfy the following constraint as well:
\begin{align}\label{sok 3'}
    \boldsymbol{s_2} - \boldsymbol{s_3} = \frac{1}{2} \,.
\end{align}
This leads to the following result:
\begin{align}\label{sok 4}
    & \left( Q_0 - 2 \right) K_a = \left[ \frac{1}{3} - \boldsymbol{s_2}, \boldsymbol{s_2}, \boldsymbol{s_2} - \frac{1}{2} \right] \nonumber\\
    & \qquad\qquad = \left( \frac{1}{3} - \boldsymbol{s_2} \right) \left(3x^{}_a + D_{1a} \right) \left( Q_0 - 2 \right) \bar\partial\cdot K \nonumber\\ 
    & \qquad\qquad + \boldsymbol{s_2} \left( Q_0 - 2 \right) x^{}_a \bar\partial\cdot K \nonumber\\ 
    & \qquad\qquad + \left( \boldsymbol{s_2} - \frac{1}{2} \right) \left( Q_0 - 2 \right) D_{1a} \bar\partial\cdot K \,.
\end{align}
After performing a series of straightforward computations (utilizing Eqs. \eqref{xxxx} and \eqref{dddd}) and rearranging the above result, it becomes evident that all terms involving $\boldsymbol{s_2}$ cancel out, leaving us with:
\begin{align}
    \left( Q_0 - 2 \right) K_a = - 2x^{}_a \bar\partial\cdot K - \frac{1}{6} D_{1a} \left( Q_0 + 4 \right) \bar\partial\cdot K \,,
\end{align}
or equivalently, employing Eq. \eqref{Q1vec}, with Eq. \eqref{sok 5}. Recall that Eq. \eqref{sok'} accompanies this equation.

At this point, a critical question naturally arises: Does Eq. \eqref{sok 5} fully capture the degrees of freedom inherent in the original equation \eqref{sok}? In other words, given the commutative nature of the operators $Q_0$ and $\left( Q_0 - 2 \right)$, instead of starting with Eq. \eqref{sok} or its counterpart \eqref{sok 1}, we could consider the following equation:
\begin{align}\label{fert}
    \left( Q_0 - 2 \right) Q_0 K_a = \left[-\frac{4}{3}, 0, 0 \right] \,.
\end{align}
The question then becomes: What would the result for $Q_0 K_a$ be if we followed a similar procedure to the one discussed above?

To address this question, we consider the three-dimensional space $\mathfrak{S}^\prime$:
\begin{align}
    \mathfrak{S}^\prime \ni [ \boldsymbol{s^\prime_1}, \boldsymbol{s^\prime_2}, \boldsymbol{s^\prime_3} ] =\, &\boldsymbol{s^\prime_1} \left[\left(3x^{}_a + D_{1a} \right) \left( Q_0 - 2 \right) \bar\partial\cdot K\right] \nonumber\\ 
    +\, & \boldsymbol{s^\prime_2} \left[Q_0 x^{}_a \bar\partial\cdot K\right] \nonumber\\
    +\, & \boldsymbol{s^\prime_3} \left[Q_0 D_{1a} \bar\partial\cdot K\right] .
\end{align}
Under the action of $\left(Q_0 - 2\right)$, we have:
\begin{align*}
    \left( Q_0 - 2 \right) \left[\left(3x^{}_a + D_{1a} \right) \left( Q_0 - 2 \right) \bar\partial\cdot K\right] =&\, [ -6, 0, 0 ]\,,\\
    \left( Q_0 - 2 \right) \left[ Q_0 x^{}_a \bar\partial\cdot K\right] =&\, [ -4, 0, 0 ]\,, \\
    \left( Q_0 - 2 \right) \left[ Q_0 D_{1a} \bar\partial\cdot K\right] =&\, [ 0, 0, 0 ]\,.
\end{align*}
Considering Eq. \eqref{fert}, the expansion of $Q_0 K_a$ within this space reads as:
\begin{align}\label{siri}
    Q_0 K_a = [ \boldsymbol{s^\prime_1}, \boldsymbol{s^\prime_2}, \boldsymbol{s^\prime_3} ] \,,
\end{align}
provided that:
\begin{eqnarray}
\left(\begin{array}{cccccccc}
-6 & -4 & 0 \\
0 & 0 & 0 \\
0 & 0 & 0
\end{array}\right)
\left(\begin{array}{cccccccc}
\boldsymbol{s^\prime_1} \\
\boldsymbol{s^\prime_2} \\
\boldsymbol{s^\prime_3}
\end{array}\right) =
\left(\begin{array}{cccccccc}
-{4}/{3} \\
0 \\
0
\end{array}\right)\,,
\end{eqnarray}
which implies that:
\begin{align}
    3\boldsymbol{s^\prime_1} + 2\boldsymbol{s^\prime_2} = \frac{2}{3}\,.
\end{align}
However, in this case, unlike the previous situation, multiplying both sides of Eq. \eqref{siri} by $x^a_{}$ and/or $\bar{\partial}^a_{}$ reveals that the solution is \underline{not} consistent. Therefore, we conclude that Eq. \eqref{sok 5} represents the only viable reduction of the original equation \eqref{sok}, and thus fully captures the degrees of freedom inherent in \eqref{sok}. This conclusion is further supported by the fact that Eq. \eqref{sok 5} remains invariant under the purely conformal transformation described in Eq. \eqref{sadafi}.

\begin{widetext}
\section{Derivation and detailed calculations for Eqs. \eqref{pmg-ci} and \eqref{ttGsolu2}} \label{App B}
Let us start with the traceless part, which satisfies Eq. \eqref{tt}, subject to the constraint \eqref{56}. Following a similar approach as outlined in the vector case (see Appendix \ref{App A}), we define a ten-dimensional space $\mathfrak{E}$, which is spanned by the linear combinations of the following set of ten fundamental functions:
\begin{align}
    \mathfrak{E} \ni [ \boldsymbol{e_1}, \,\ldots \,, \boldsymbol{e_{10}} ] =\, & \boldsymbol{e_1} \left[{\cal{S}}x^{}_a ( 3x^{}_b + D_{1b} ) \left( Q_0 - 2 \right) \bar\partial\cdot\bar\partial\cdot{\cal{K}}^0 \right] + \boldsymbol{e_2} \left[{\cal{S}}D^{}_{1a} ( 3x^{}_b + D_{1b} ) \left( Q_0 - 2 \right) \bar\partial\cdot\bar\partial\cdot{\cal{K}}^0 \right] \nonumber\\
    +\,& \boldsymbol{e_3} \left[{\cal{S}} ( 3x^{}_a + D_{1a} ) \left( Q_0 - 2 \right) ( 3x^{}_b + D_{1b} )\; \bar\partial\cdot\bar\partial\cdot{\cal{K}}^0 \right] \nonumber\\
    +\,& \boldsymbol{e_4} \left[{\cal{S}} ( 3x^{}_a + D_{1a} ) \left( Q_0 - 2 \right) x^{}_b \bar\partial\cdot\bar\partial\cdot{\cal{K}}^0 \right] \nonumber\\
    +\,& \boldsymbol{e_5} \left[{\cal{S}} \left( Q_0 - 2 \right) 6 x^{}_a \bar\partial\cdot{\cal{K}}^0_{\cdot b} \right] + \boldsymbol{e_6} \left[{\cal{S}} \left( Q_0 - 2 \right) 6 D_{1a} \bar\partial\cdot{\cal{K}}^0_{\cdot b} \right] \nonumber\\
    +\,& \boldsymbol{e_7} \left[{\cal{S}} \left( Q_0 - 2 \right) x^{}_a x^{}_b \, \bar\partial\cdot\bar\partial\cdot{\cal{K}}^0 \right] + \boldsymbol{e_8} \left[{\cal{S}} \left( Q_0 - 2 \right) x^{}_a D_{1b} \, \bar\partial\cdot\bar\partial\cdot{\cal{K}}^0 \right] \nonumber\\
    +\,& \boldsymbol{e_9} \left[{\cal{S}} \left( Q_0 - 2 \right) D_{1a} x^{}_b \, \bar\partial\cdot\bar\partial\cdot{\cal{K}}^0 \right] + \boldsymbol{e_{10}} \left[{\cal{S}} \left( Q_0 - 2 \right) D_{1a} D_{1b} \, \bar\partial\cdot\bar\partial\cdot{\cal{K}}^0 \right] \,.
\end{align}
This space remains invariant under the action of $Q_0$:
\begin{align}
    Q_0 \left[{\cal{S}} x^{}_a ( 3x^{}_b + D_{1b} ) \left( Q_0 - 2 \right) \bar\partial\cdot\bar\partial\cdot{\cal{K}}^0 \right] =&\, [ -8, -2, 0, \,\ldots\,, 0] \,,\\
    Q_0 \left[{\cal{S}} D^{}_{1a} ( 3x^{}_b + D_{1b} ) \left( Q_0 - 2 \right) \bar\partial\cdot\bar\partial\cdot{\cal{K}}^0 \right] =&\, [ -8, -2, 0, \,\ldots\,, 0] \,,\\ 
    Q_0 \left[{\cal{S}} ( 3x^{}_a + D_{1a} ) \left( Q_0 - 2 \right) ( 3x^{}_b + D_{1b} )\; \bar\partial\cdot\bar\partial\cdot{\cal{K}}^0 \right] =&\, [ -60, -12, -4, 0, \,\ldots\,, 0] \,,\\
    Q_0 \left[{\cal{S}} ( 3x^{}_a + D_{1a} ) \left( Q_0 - 2 \right) x^{}_b \bar\partial\cdot\bar\partial\cdot{\cal{K}}^0 \right] =&\, [ -20, -4, 0, -4, 0, \,\ldots\,, 0] \,, \\
    Q_0 \left[{\cal{S}} \left( Q_0 - 2 \right) 6 x^{}_a \bar\partial\cdot{\cal{K}}^0_{\cdot b} \right] =&\, [-32, -8, 12, -36, \,\ldots\,, 0] \,,\\
    Q_0 \left[{\cal{S}} \left( Q_0 - 2 \right) 6 D_{1a} \bar\partial\cdot{\cal{K}}^0_{\cdot b} \right] =&\, [-32, -8, 0, 0, \,\ldots\,, 0] \,,\\    
    Q_0 \left[{\cal{S}} \left( Q_0 - 2 \right) x^{}_a x^{}_b \, \bar\partial\cdot\bar\partial\cdot{\cal{K}}^0 \right] =&\, [-4, 0, 0, -4, 0, \,\ldots\,, 0] \,,\\
    Q_0 \left[{\cal{S}} \left( Q_0 - 2 \right) x^{}_a D_{1b} \, \bar\partial\cdot\bar\partial\cdot{\cal{K}}^0 \right] =&\, [0, 0, -4, 12, 0, \,\ldots\,, 0] \,,\\
    Q_0 \left[{\cal{S}} \left( Q_0 - 2 \right) D_{1a} x^{}_b \, \bar\partial\cdot\bar\partial\cdot{\cal{K}}^0 \right] =&\, [-16, -4, 0, \,\ldots\,, 0] \,,\\
    Q_0 \left[{\cal{S}} \left( Q_0 - 2 \right) D_{1a} D_{1b} \, \bar\partial\cdot\bar\partial\cdot{\cal{K}}^0 \right] =&\, [0, \,\ldots\,, 0] \,.
\end{align}

Then, Eq. \eqref{tt} can be rewritten as:
\begin{align}\label{tt'}
    Q_0 \left( Q_0 - 2 \right) {\cal{K}}^{0}_{ab} = \frac{1}{6} \left[ -7, -2, 3, -8, 0, \,\ldots\,, 0 \right] \,.
\end{align}
It can be reduced to:
\begin{align}\label{tt''}
    \left( Q_0 - 2 \right) {\cal{K}}^{0}_{ab} = \left[ \boldsymbol{e_1}, \,\ldots\, , \boldsymbol{e_{10}}\right] \,,
\end{align}
provided that:
\begin{eqnarray}
\left(\begin{array}{cccccccccc}
-8 & -8 & -60 & -20 & -32 & -32 & -4 & 0 & -16 & 0 \\
-2 & -2 & -12 & -4 & -8 & -8 & 0 & 0 & -4 & 0 \\
0 & 0 & -4 & 0 & 12 & 0 & 0 & -4 & 0 & 0 \\
0 & 0 & 0 & -4 & -36 & 0 & -4 & 12 & 0 & 0 \\
0 & 0 & 0 & 0 & 0 & 0 & 0 & 0 & 0 & 0 \\
\vdots & \vdots & \vdots & \vdots & \vdots & \vdots & \vdots & \vdots & \vdots & \vdots \\
0 & 0 & 0 & 0 & 0 & 0 & 0 & 0 & 0 & 0 
\end{array}\right)
\left(\begin{array}{cccccccccc}
\boldsymbol{e_1} \\
\boldsymbol{e_2} \\
\boldsymbol{e_3} \\
\boldsymbol{e_4} \\
\boldsymbol{e_5} \\
\vdots \\
\boldsymbol{e_{10}}
\end{array}\right) = \frac{1}{6}
\left(\begin{array}{cccccccc}
-7 \\
-2 \\
3 \\
-8 \\
0 \\
\vdots \\
0
\end{array}\right)\,,
\end{eqnarray}
which gives rise to the following three independent constraints:
\begin{align}
    \label{abbas1} & \boldsymbol{e_4} + 9\boldsymbol{e_5} + \boldsymbol{e_7} - 3\boldsymbol{e_8} = \frac{1}{3}\,, \\
    \label{abbas2} & \boldsymbol{e_3} - 3\boldsymbol{e_5} + \boldsymbol{e_8} = -\frac{1}{8}\,, \\
    \label{abbas3} & \boldsymbol{e_1} + \boldsymbol{e_2} + 6\boldsymbol{e_3} + 2\boldsymbol{e_4} + 4\boldsymbol{e_5} + 4\boldsymbol{e_6} + 2\boldsymbol{e_9} = \frac{1}{6}\,.
\end{align}

Next, multiplying both sides of Eq. \eqref{tt''} by $x^{a}_{}$ shows that the consistent solution must satisfy:
\begin{align}
    & \boldsymbol{e_5} - \boldsymbol{e_6} = \frac{1}{18}\,,\\
    & 6\boldsymbol{e_1} + 30\boldsymbol{e_3} + 6\boldsymbol{e_4} + 2\boldsymbol{e_7} + 4\boldsymbol{e_8} + 4\boldsymbol{e_9} = 0 \,,\\
    & \boldsymbol{e_1} + 4\boldsymbol{e_2} + 9\boldsymbol{e_3} + \boldsymbol{e_4} - \boldsymbol{e_5} - 3\boldsymbol{e_6} + \boldsymbol{e_8} + \boldsymbol{e_9} + 9\boldsymbol{e_{10}} = 0\,,\\
    & 12\boldsymbol{e_1} + 108\boldsymbol{e_3} + 36\boldsymbol{e_4} + 24\boldsymbol{e_5} + 48\boldsymbol{e_6} + 20\boldsymbol{e_7} + 16\boldsymbol{e_9} = 0\,,\\
    & 2\boldsymbol{e_1} + 8\boldsymbol{e_2} + 42\boldsymbol{e_3} + 14\boldsymbol{e_4} + 4\boldsymbol{e_5} + 24\boldsymbol{e_6} + 4\boldsymbol{e_7} + 6\boldsymbol{e_8} + 10\boldsymbol{e_9} - 6\boldsymbol{e_{10}} = 0\,.
\end{align}
On the other hand, multiplying both sides of Eq. \eqref{tt''} by $\eta^{ab}_{}$ reveals that the consistent solution must also satisfy:
\begin{align}
    -3\boldsymbol{e_1} + 12\boldsymbol{e_2} + 9\boldsymbol{e_3} + 3\boldsymbol{e_4} + 6\boldsymbol{e_6} - \boldsymbol{e_7} + 4\boldsymbol{e_9} = 0\,.
\end{align}

By utilizing the online tools outlined in \cite{online}, we obtain the following results verifying the above set of equations:
\begin{align}\label{tt'''}
    \left( Q_0 - 2 \right) {\cal{K}}^{0}_{ab} = \frac{1}{216} &\big[ \boldsymbol{e_1} = -5,\; \boldsymbol{e_2} = -1,\; \boldsymbol{e_3} = 3,\; \boldsymbol{e_4} = -18 -216\boldsymbol{e_7},\; \boldsymbol{e_5} = 12, \nonumber\\
    &\hspace{0.66cm} \boldsymbol{e_6} = 0,\; 216\boldsymbol{e_7},\; \boldsymbol{e_8} = 6,\; \boldsymbol{e_9} = 6 + 216\boldsymbol{e_7},\; \boldsymbol{e_{10}} =0 \big] \nonumber\\
    = \frac{1}{216}& \big( -5 \left[{\cal{S}}x^{}_a ( 3x^{}_b + D_{1b} ) \left( Q_0 - 2 \right) \bar\partial\cdot\bar\partial\cdot{\cal{K}}^0 \right] -\left[{\cal{S}}D^{}_{1a} ( 3x^{}_b + D_{1b} ) \left( Q_0 - 2 \right) \bar\partial\cdot\bar\partial\cdot{\cal{K}}^0 \right] \nonumber\\
    &\;\,+ 3 \left[{\cal{S}} ( 3x^{}_a + D_{1a} ) \left( Q_0 - 2 \right) ( 3x^{}_b + D_{1b} )\; \bar\partial\cdot\bar\partial\cdot{\cal{K}}^0 \right] \nonumber\\
    &\;\,+ \left(-18-216\boldsymbol{e_7}\right) \left[{\cal{S}} ( 3x^{}_a + D_{1a} ) \left( Q_0 - 2 \right) x^{}_b \bar\partial\cdot\bar\partial\cdot{\cal{K}}^0 \right] \nonumber\\
    &\;\,+ 12 \left[{\cal{S}} \left( Q_0 - 2 \right) 6 x^{}_a \bar\partial\cdot{\cal{K}}^0_{\cdot b} \right] + 216\boldsymbol{e_7} \left[{\cal{S}} \left( Q_0 - 2 \right) x^{}_a x^{}_b \, \bar\partial\cdot\bar\partial\cdot{\cal{K}}^0 \right] \nonumber\\
    &\;\,+ 6 \left[{\cal{S}} \left( Q_0 - 2 \right) x^{}_a D_{1b} \, \bar\partial\cdot\bar\partial\cdot{\cal{K}}^0 \right] + \left(6+216\boldsymbol{e_7}\right) \left[{\cal{S}} \left( Q_0 - 2 \right) D_{1a} x^{}_b \, \bar\partial\cdot\bar\partial\cdot{\cal{K}}^0 \right] \big)\,.
\end{align}
After carrying out a series of straightforward calculations (utilizing Eqs. \eqref{xxxx} and \eqref{dddd}) and reorganizing the result, it becomes clear that all terms involving $\boldsymbol{e_7}$ cancel out, leaving us with the following simple representation:
\begin{align}\label{ttsolu0}
    \left( Q_0 - 2 \right) {\cal{K}}^{0}_{ab} = -2 {\cal{S}} x^{}_a \bar\partial\cdot{\cal{K}}^0_{\cdot b} - \frac{2}{3} D_{2a} \bar\partial\cdot{\cal{K}}^0_{\cdot b} + \frac{1}{3} R^2 \theta_{ab} \bar\partial\cdot\bar\partial\cdot{\cal{K}}^0_{} + \frac{1}{108} D_{2a}D_{1b}\left( Q_0 - 2 \right) \bar\partial\cdot\bar\partial\cdot{\cal{K}}^0_{} \,,
\end{align}
or alternatively, utilizing Eq. \eqref{Q1ten}, with:
\begin{align}\label{ttsolu2}
    \left( Q_2 + 4 \right) {\cal{K}}^{0}_{ab} = - \frac{2}{3} D_{2a} \bar\partial\cdot{\cal{K}}^0_{\cdot b} + \frac{1}{3} R^2 \theta_{ab} \bar\partial\cdot\bar\partial\cdot{\cal{K}}^0_{} + \frac{1}{108} D_{2a}D_{1b}\left( Q_0 - 2 \right) \bar\partial\cdot\bar\partial\cdot{\cal{K}}^0_{}\,.
\end{align}
Once again, by multiplying the above result by $x^a_{}$ and/or $\eta^{ab}_{}$, one can verify its consistency.

We now turn our attention to the pure-trace component, which satisfies Eq. \eqref{pt} under the constraint given by Eq. \eqref{65}. To proceed, we introduce the following eight-dimensional space, denoted as $\mathfrak{U}$:
\begin{align}
    \mathfrak{U} \ni [ \boldsymbol{u_1}, \,\ldots \,, \boldsymbol{u_8} ] =\, & \boldsymbol{u_1} \left[{\cal{S}} R^{-2} x^{}_a ( 3x^{}_b + D_{1b} ) \left( Q_0 - 2 \right) {\cal{K}}^\prime \right] + \boldsymbol{u_2} \left[{\cal{S}} R^{-2} D^{}_{1a} ( 3x^{}_b + D_{1b} ) \left( Q_0 - 2 \right) {\cal{K}}^\prime \right] \nonumber\\
    +\,& \boldsymbol{u_3} \left[{\cal{S}} R^{-2} ( 3x^{}_a + D_{1a} ) \left( Q_0 - 2 \right) ( 3x^{}_b + D_{1b} )\; {\cal{K}}^\prime \right] + \boldsymbol{u_4} \left[{\cal{S}} R^{-2} ( 3x^{}_a + D_{1a} ) \left( Q_0 - 2 \right) x^{}_b {\cal{K}}^\prime \right] \nonumber\\
    +\,& \boldsymbol{u_5} \left[{\cal{S}} R^{-2} \left( Q_0 - 2 \right) x^{}_a x^{}_b \, {\cal{K}}^\prime \right] + \boldsymbol{u_6} \left[{\cal{S}} R^{-2} \left( Q_0 - 2 \right) x^{}_a D_{1b} \, {\cal{K}}^\prime \right] \nonumber\\
    +\,& \boldsymbol{u_7} \left[{\cal{S}} R^{-2} \left( Q_0 - 2 \right) D_{1a} x^{}_b \, {\cal{K}}^\prime \right] + \boldsymbol{u_8} \left[{\cal{S}} R^{-2} \left( Q_0 - 2 \right) D_{1a} D_{1b} \, {\cal{K}}^\prime \right] \,.
\end{align}
This space remains invariant under the action of $Q_0$:
\begin{align}
    Q_0 \left[{\cal{S}} R^{-2} x^{}_a ( 3x^{}_b + D_{1b} ) \left( Q_0 - 2 \right) {\cal{K}}^\prime \right] =&\, [ -8, -2, 0, \,\ldots\,, 0] \,,\\
    Q_0 \left[{\cal{S}} R^{-2} D^{}_{1a} ( 3x^{}_b + D_{1b} ) \left( Q_0 - 2 \right) {\cal{K}}^\prime \right] =&\, [ -8, -2, 0, \,\ldots\,, 0] \,,\\ 
    Q_0 \left[{\cal{S}} R^{-2} ( 3x^{}_a + D_{1a} ) \left( Q_0 - 2 \right) ( 3x^{}_b + D_{1b} )\; {\cal{K}}^\prime \right] =&\, [ -60, -12, -4, 0, \,\ldots\,, 0] \,,\\
    Q_0 \left[{\cal{S}} R^{-2} ( 3x^{}_a + D_{1a} ) \left( Q_0 - 2 \right) x^{}_b {\cal{K}}^\prime \right] =&\, [ -20, -4, 0, -4, 0, \,\ldots\,, 0] \,, \\
    Q_0 \left[{\cal{S}} R^{-2} \left( Q_0 - 2 \right) x^{}_a x^{}_b \, {\cal{K}}^\prime \right] =&\, [-4, 0, 0, -4, 0, \,\ldots\,, 0] \,,\\
    Q_0 \left[{\cal{S}} R^{-2} \left( Q_0 - 2 \right) x^{}_a D_{1b} \, {\cal{K}}^\prime \right] =&\, [0, 0, -4, 12, 0, \,\ldots\,, 0] \,,\\
    Q_0 \left[{\cal{S}} R^{-2} \left( Q_0 - 2 \right) D_{1a} x^{}_b \, {\cal{K}}^\prime \right] =&\, [-16, -4, 0, \,\ldots\,, 0] \,,\\
    Q_0 \left[{\cal{S}} R^{-2} \left( Q_0 - 2 \right) D_{1a} D_{1b} \, {\cal{K}}^\prime \right] =&\, [0, \,\ldots\,, 0] \,.
\end{align}

Subsequently, Eq. \eqref{pt} can be rewritten as:
\begin{align}\label{pt'}
    Q_0 \left( Q_0 - 2 \right) \theta_{ab} {\cal{K}}^\prime = \left[ -2, 0, 0, -2, 0, \,\ldots\,, 0 \right] \,.
\end{align}
It can be reduced to:
\begin{align}\label{pt''}
    \left( Q_0 - 2 \right) \theta_{ab} {\cal{K}}^\prime = \left[ \boldsymbol{u_1}, \,\ldots\, , \boldsymbol{u_8}\right] \,,
\end{align}
such that:
\begin{eqnarray}
\left(\begin{array}{cccccccccc}
-8 & -8 & -60 & -20 & -4 & 0 & -16 & 0 \\
-2 & -2 & -12 & -4 & 0 & 0 & -4 & 0 \\
0 & 0 & -4 & 0 & 0 & -4 & 0 & 0 \\
0 & 0 & 0 & -4 & -4 & 12 & 0 & 0 \\
0 & 0 & 0 & 0 & 0 & 0 & 0 & 0 \\
\vdots & \vdots & \vdots & \vdots & \vdots & \vdots & \vdots & \vdots \\
0 & 0 & 0 & 0 & 0 & 0 & 0 & 0 
\end{array}\right)
\left(\begin{array}{cccccccccc}
\boldsymbol{u_1} \\
\boldsymbol{u_2} \\
\boldsymbol{u_3} \\
\boldsymbol{u_4} \\
\boldsymbol{u_5} \\
\vdots \\
\boldsymbol{u_8}
\end{array}\right) = 
\left(\begin{array}{cccccccc}
-2 \\
0 \\
0 \\
-2 \\
0 \\
\vdots \\
0
\end{array}\right)\,.
\end{eqnarray}
The latter implies that:
\begin{align}
    \label{abbas11} & \boldsymbol{u_3} + \boldsymbol{u_6} = 0\,, \\
    \label{abbas22} & 2\boldsymbol{u_4} + 2\boldsymbol{u_5} - 6\boldsymbol{u_6} = 1\,, \\
    \label{abbas33} & \boldsymbol{u_1} + \boldsymbol{u_2} + 6\boldsymbol{u_3} + 2\boldsymbol{u_4} + 2\boldsymbol{u_7} = 0\,.
\end{align}

Multiplying both sides of Eq. \eqref{pt''} by $x^a_{}$ reveals that the consistent solution must satisfy:
\begin{align}
    & 6\boldsymbol{u_1} + 30\boldsymbol{u_3} + 6\boldsymbol{u_4} + 2\boldsymbol{u_5} + 4\boldsymbol{u_6} + 4\boldsymbol{u_7} = 0 \,,\\
    & \boldsymbol{u_1} + 4\boldsymbol{u_2} + 9\boldsymbol{u_3} + \boldsymbol{u_4} + \boldsymbol{u_6} + \boldsymbol{u_7} + 9\boldsymbol{u_8} = 0\,,\\
    & 12\boldsymbol{u_1} + 108\boldsymbol{u_3} + 36\boldsymbol{u_4} + 20\boldsymbol{u_5} + 16\boldsymbol{u_7} = 8\,,\\
    & 2\boldsymbol{u_1} + 8\boldsymbol{u_2} + 42\boldsymbol{u_3} + 14\boldsymbol{u_4} + 4\boldsymbol{u_5} + 6\boldsymbol{u_6} + 10\boldsymbol{u_7} - 6\boldsymbol{u_8} = 2\,.
\end{align}
Alternatively, multiplying both sides of Eq. \eqref{pt''} by $\eta^{ab}_{}$ shows that the consistent solution must also satisfy:
\begin{align}
    -3\boldsymbol{u_1} + 12\boldsymbol{u_2} + 9\boldsymbol{u_3} + 3\boldsymbol{u_4} - \boldsymbol{u_5} + 4\boldsymbol{u_7} = 2\,.
\end{align}

Again, by using the online tools mentioned in \cite{online}, we obtain:
\begin{align}\label{pt'''}
    \left( Q_0 - 2 \right) \theta_{ab} {\cal{K}}^\prime = \frac{1}{2} &\big[ \boldsymbol{u_1} = -3,\; \boldsymbol{u_2} = -1,\; \boldsymbol{u_3} = 1,\; \boldsymbol{u_4} = -2 - 2\boldsymbol{u_5},\; 2\boldsymbol{u_5},\; \boldsymbol{u_6} = -1,\; \boldsymbol{u_7} = 1 + 2\boldsymbol{u_5},\; \boldsymbol{u_8} = 0 \big] \nonumber\\
    = \frac{1}{2}& \big( -3 \left[{\cal{S}} R^{-2} x^{}_a ( 3x^{}_b + D_{1b} ) \left( Q_0 - 2 \right) {\cal{K}}^\prime \right] -\left[{\cal{S}} R^{-2} D^{}_{1a} ( 3x^{}_b + D_{1b} ) \left( Q_0 - 2 \right) {\cal{K}}^\prime \right] \nonumber\\
    &\;\,+ \left[{\cal{S}} R^{-2} ( 3x^{}_a + D_{1a} ) \left( Q_0 - 2 \right) ( 3x^{}_b + D_{1b} )\; {\cal{K}}^\prime \right] \nonumber\\
    &\;\,+ \left(-2-2\boldsymbol{u_5}\right) \left[{\cal{S}} R^{-2} ( 3x^{}_a + D_{1a} ) \left( Q_0 - 2 \right) x^{}_b {\cal{K}}^\prime \right] + 2\boldsymbol{u_5} \left[{\cal{S}} R^{-2} \left( Q_0 - 2 \right) x^{}_a x^{}_b \, {\cal{K}}^\prime \right] \nonumber\\
    &\;\,- \left[{\cal{S}} R^{-2} \left( Q_0 - 2 \right) x^{}_a D_{1b} \, {\cal{K}}^\prime \right] + \left(1+2\boldsymbol{u_5}\right) \left[{\cal{S}} R^{-2} \left( Q_0 - 2 \right) D_{1a} x^{}_b \, {\cal{K}}^\prime \right] \big)\,.
\end{align}
After some simple calculations, we arrive at:
\begin{align}\label{ptsolu1}
    \left( Q_0 - 2 \right) \theta_{ab} {\cal{K}}^\prime = \theta_{ab} Q_0 {\cal{K}}^\prime - 2R^{-2}D_{2a} x^{}_b {\cal{K}}^\prime - 12R^{-2}x^{}_ax^{}_b {\cal{K}}^\prime \,,
\end{align}
or equivalently, employing Eq. \eqref{Q1ten}, at:
\begin{align}\label{ptsolu2}
    \left( Q_2 + 4 \right) \theta_{ab} {\cal{K}}^\prime = \theta_{ab} \left( Q_0 + 4 \right) {\cal{K}}^\prime \,,
\end{align}
which is independent of the coefficient $\boldsymbol{u_5}$. Note that: (i) In deriving the latter equation, we have used the relation $\bar\partial\cdot\theta{\cal{K}}^\prime = \bar\partial{\cal{K}}^\prime + 4R^{-2} x {\cal{K}}^\prime$. Additionally, for future reference, we may consider the following identity $\bar\partial\cdot\bar\partial\cdot\theta{\cal{K}}^\prime = R^{-2} \left(-Q_0 + 16 \right){\cal{K}}^\prime$. (ii) The above result can be interpreted as the commutation relation between the operators $Q_2$ and $\theta$, though it is somewhat trivial. However, this procedure ensures that Eq. \eqref{pt} does not yield any significant result when reduced to a quadratic form, as we did for the traceless part.

Now, combining the traceless and pure-trace components, given by Eqs. \eqref{ttsolu2} and \eqref{ptsolu2} respectively, a straightforward calculation produces Eq. \eqref{pmg-ci}.

Similar to the vector case (see Appendix \ref{App A}), due to the commutative nature of the operators $Q_0$ and $\left( Q_0 - 2 \right)$, we can consider the following equation instead of starting with Eq. \eqref{tt}:
\begin{align}
    \left( Q_0 - 2 \right) Q_0 {\cal{K}}^0_{ab} =&\, \frac{1}{6} {\cal{S}} \big[ -2 D_{1a} (3x^{}_b + D_{1b}) \left( Q_0 - 2 \right) + 3 (3x^{}_a + D_{1a}) \left( Q_0 - 2 \right) (3x^{}_b + D_{1b}) \nonumber\\
    &\quad\quad - 7 x^{}_a(3x^{}_b + D_{1b}) \left( Q_0 - 2 \right) - 8 (3x^{}_a + D_{1a}) \left( Q_0 - 2 \right) x^{}_b \big] \bar\partial\cdot\bar\partial\cdot{\cal{K}}^0_{}
    \label{ttG}\,, 
\end{align}
of course, with the constraints \eqref{56} and \eqref{65}.

In this case, we alternatively begin with the following ten-dimensional space $\mathfrak{E}^\prime$:
\begin{align}
    \mathfrak{E}^\prime \ni [ \boldsymbol{e^\prime_1}, \,\ldots \,, \boldsymbol{e^\prime_{10}} ] =\, & \boldsymbol{e^\prime_1} \left[{\cal{S}}x^{}_a ( 3x^{}_b + D_{1b} ) \left( Q_0 - 2 \right) \bar\partial\cdot\bar\partial\cdot{\cal{K}}^0 \right] + \boldsymbol{e^\prime_2} \left[{\cal{S}}D^{}_{1a} ( 3x^{}_b + D_{1b} ) \left( Q_0 - 2 \right) \bar\partial\cdot\bar\partial\cdot{\cal{K}}^0 \right] \nonumber\\
    +\,& \boldsymbol{e^\prime_3} \left[{\cal{S}} ( 3x^{}_a + D_{1a} ) \left( Q_0 - 2 \right) ( 3x^{}_b + D_{1b} )\; \bar\partial\cdot\bar\partial\cdot{\cal{K}}^0 \right] \nonumber\\
    +\,& \boldsymbol{e^\prime_4} \left[{\cal{S}} ( 3x^{}_a + D_{1a} ) \left( Q_0 - 2 \right) x^{}_b \bar\partial\cdot\bar\partial\cdot{\cal{K}}^0 \right] + \boldsymbol{e^\prime_5} \left[{\cal{S}} Q_0 6 x^{}_a \bar\partial\cdot{\cal{K}}^0_{\cdot b} \right] \nonumber\\
    +\,& \boldsymbol{e^\prime_6} \left[{\cal{S}} Q_0 6 D_{1a} \bar\partial\cdot{\cal{K}}^0_{\cdot b} \right] + \boldsymbol{e^\prime_7} \left[{\cal{S}} Q_0 x^{}_a x^{}_b \, \bar\partial\cdot\bar\partial\cdot{\cal{K}}^0 \right] + \boldsymbol{e^\prime_8} \left[{\cal{S}} Q_0 x^{}_a D_{1b} \, \bar\partial\cdot\bar\partial\cdot{\cal{K}}^0 \right] \nonumber\\
    +\,& \boldsymbol{e^\prime_9} \left[{\cal{S}} Q_0 D_{1a} x^{}_b \, \bar\partial\cdot\bar\partial\cdot{\cal{K}}^0 \right] + \boldsymbol{e^\prime_{10}} \left[{\cal{S}} Q_0 D_{1a} D_{1b} \, \bar\partial\cdot\bar\partial\cdot{\cal{K}}^0 \right] \,.
\end{align}
This space remains invariant under the action of $\left(Q_0 - 2\right)$:
\begin{align}
    \left(Q_0 - 2\right) \left[{\cal{S}} x^{}_a ( 3x^{}_b + D_{1b} ) \left( Q_0 - 2 \right) \bar\partial\cdot\bar\partial\cdot{\cal{K}}^0 \right] =&\, [ -10, -2, 0, \,\ldots\,, 0] \,,\\
    \left(Q_0 - 2\right) \left[{\cal{S}} D^{}_{1a} ( 3x^{}_b + D_{1b} ) \left( Q_0 - 2 \right) \bar\partial\cdot\bar\partial\cdot{\cal{K}}^0 \right] =&\, [ -8, -4, 0, \,\ldots\,, 0] \,,\\ 
    \left(Q_0 - 2\right) \left[{\cal{S}} ( 3x^{}_a + D_{1a} ) \left( Q_0 - 2 \right) ( 3x^{}_b + D_{1b} )\; \bar\partial\cdot\bar\partial\cdot{\cal{K}}^0 \right] =&\, [ -60, -12, -6, 0, \,\ldots\,, 0] \,,\\
    \left(Q_0 - 2\right) \left[{\cal{S}} ( 3x^{}_a + D_{1a} ) \left( Q_0 - 2 \right) x^{}_b \bar\partial\cdot\bar\partial\cdot{\cal{K}}^0 \right] =&\, [ -20, -4, 0, -6, 0, \,\ldots\,, 0] \,, \\    \left(Q_0 - 2\right) \left[{\cal{S}} Q_0 6 x^{}_a \bar\partial\cdot{\cal{K}}^0_{\cdot b} \right] =&\, [-32, -8, 12, -36, \,\ldots\,, 0] \,,\\
    \left(Q_0 - 2\right) \left[{\cal{S}} Q_0 6 D_{1a} \bar\partial\cdot{\cal{K}}^0_{\cdot b} \right] =&\, [-32, -8, 0, 0, \,\ldots\,, 0] \,,\\    
    \left(Q_0 - 2\right) \left[{\cal{S}} Q_0 x^{}_a x^{}_b \, \bar\partial\cdot\bar\partial\cdot{\cal{K}}^0 \right] =&\, [-4, 0, 0, -4, 0, \,\ldots\,, 0] \,,\\
    \left(Q_0 - 2\right) \left[{\cal{S}} Q_0 x^{}_a D_{1b} \, \bar\partial\cdot\bar\partial\cdot{\cal{K}}^0 \right] =&\, [0, 0, -4, 12, 0, \,\ldots\,, 0] \,,\\
    \left(Q_0 - 2\right) \left[{\cal{S}} Q_0 D_{1a} x^{}_b \, \bar\partial\cdot\bar\partial\cdot{\cal{K}}^0 \right] =&\, [-16, -4, 0, \,\ldots\,, 0] \,,\\
    \left(Q_0 - 2\right) \left[{\cal{S}} Q_0 D_{1a} D_{1b} \, \bar\partial\cdot\bar\partial\cdot{\cal{K}}^0 \right] =&\, [0, \,\ldots\,, 0] \,.
\end{align}

The assumed Eq. \eqref{ttG} then takes the form:
\begin{align}\label{ttG'}
    \left( Q_0 - 2 \right) Q_0 {\cal{K}}^{0}_{ab} = \frac{1}{6} \left[ -7, -2, 3, -8, 0, \,\ldots\,, 0 \right] \,.
\end{align}
Subsequently, we can write:
\begin{align}\label{ttG''}
    Q_0 {\cal{K}}^{0}_{ab} = \left[ \boldsymbol{e^\prime_1}, \,\ldots\, , \boldsymbol{e^\prime_{10}}\right] \,,
\end{align}
provided that:
\begin{eqnarray}
\left(\begin{array}{cccccccccc}
-10 & -8 & -60 & -20 & -32 & -32 & -4 & 0 & -16 & 0 \\
-2 & -4 & -12 & -4 & -8 & -8 & 0 & 0 & -4 & 0 \\
0 & 0 & -6 & 0 & 12 & 0 & 0 & -4 & 0 & 0 \\
0 & 0 & 0 & -6 & -36 & 0 & -4 & 12 & 0 & 0 \\
0 & 0 & 0 & 0 & 0 & 0 & 0 & 0 & 0 & 0 \\
\vdots & \vdots & \vdots & \vdots & \vdots & \vdots & \vdots & \vdots & \vdots & \vdots \\
0 & 0 & 0 & 0 & 0 & 0 & 0 & 0 & 0 & 0 
\end{array}\right)
\left(\begin{array}{cccccccccc}
\boldsymbol{e^\prime_1} \\
\boldsymbol{e^\prime_2} \\
\boldsymbol{e^\prime_3} \\
\boldsymbol{e^\prime_4} \\
\boldsymbol{e^\prime_5} \\
\vdots \\
\boldsymbol{e_{10}}
\end{array}\right) = \frac{1}{6}
\left(\begin{array}{cccccccc}
-7 \\
-2 \\
3 \\
-8 \\
0 \\
\vdots \\
0
\end{array}\right)\,.
\end{eqnarray}
This immediately results in:
\begin{align}
    \label{abbasG1} & 6\boldsymbol{e^\prime_4} + 36\boldsymbol{e^\prime_5} + 4\boldsymbol{e^\prime_7} - 12\boldsymbol{e^\prime_8} = \frac{4}{3}\,, \\
    \label{abbasG2} & 6\boldsymbol{e^\prime_3} - 12\boldsymbol{e^\prime_5} + 4\boldsymbol{e^\prime_8} = -\frac{1}{2}\,, \\
    \label{abbasG3} & 2\boldsymbol{e^\prime_1} + 4\boldsymbol{e^\prime_2} + 12\boldsymbol{e^\prime_3} + 4\boldsymbol{e^\prime_4} + 8\boldsymbol{e^\prime_5} + 8\boldsymbol{e^\prime_6} + 4\boldsymbol{e^\prime_9} = \frac{1}{3} \,, \\
    \label{abbasG4} & 10\boldsymbol{e^\prime_1} + 8\boldsymbol{e^\prime_2} + 60\boldsymbol{e^\prime_3} + 20\boldsymbol{e^\prime_4} + 32\boldsymbol{e^\prime_5} + 32\boldsymbol{e^\prime_6} + 4\boldsymbol{e^\prime_7} + 16\boldsymbol{e^\prime_9} = \frac{7}{6}\,.
\end{align}

Next, multiplying both sides of Eq. \eqref{ttG''} by $x^{a}_{}$ shows that the consistent solution must satisfy:
\begin{align}
    & \boldsymbol{e^\prime_5} - 2\boldsymbol{e^\prime_6} = \frac{1}{12}\,,\\
    & 6\boldsymbol{e^\prime_1} + 30\boldsymbol{e^\prime_3} + 6\boldsymbol{e^\prime_4} + 2\boldsymbol{e^\prime_7} + 4\boldsymbol{e^\prime_8} + 4\boldsymbol{e^\prime_9} = 0 \,,\\
    & \boldsymbol{e^\prime_1} + 4\boldsymbol{e^\prime_2} + 9\boldsymbol{e^\prime_3} + \boldsymbol{e^\prime_4} - \boldsymbol{e^\prime_5} - 3\boldsymbol{e^\prime_6} + \boldsymbol{e^\prime_8} + \boldsymbol{e^\prime_9} + 9\boldsymbol{e^\prime_{10}} = 0\,,\\
    & 12\boldsymbol{e^\prime_1} + 108\boldsymbol{e^\prime_3} + 36\boldsymbol{e^\prime_4} + 24\boldsymbol{e^\prime_5} + 48\boldsymbol{e^\prime_6} + 16\boldsymbol{e^\prime_7} + 16\boldsymbol{e^\prime_9} = 0\,,\\
    & 2\boldsymbol{e^\prime_1} + 8\boldsymbol{e^\prime_2} + 42\boldsymbol{e^\prime_3} + 14\boldsymbol{e^\prime_4} + 4\boldsymbol{e^\prime_5} + 24\boldsymbol{e^\prime_6} + 4\boldsymbol{e^\prime_7} + 4\boldsymbol{e^\prime_8} + 8\boldsymbol{e^\prime_9} - 8\boldsymbol{e^\prime_{10}} = 0\,.
\end{align}
Multiplying both sides of Eq. \eqref{ttG''} by $\eta^{ab}_{}$ shows that the consistent solution must also fulfill:
\begin{align}
    -&3\boldsymbol{e^\prime_1} + 12\boldsymbol{e^\prime_2} + 9\boldsymbol{e^\prime_3} + 3\boldsymbol{e^\prime_4} = 0 \,, \\
    &6\boldsymbol{e^\prime_6} - \boldsymbol{e^\prime_7} + 4\boldsymbol{e^\prime_9} - 2\boldsymbol{e^\prime_{10}} = 0\,.
\end{align}

Again, by employing the online tools given in \cite{online}, we drive the following results satisfying the above set of equations:
\begin{align}\label{ttG'''}
    Q_0 {\cal{K}}^{0}_{ab} = \frac{1}{72} &\big[ \boldsymbol{e^\prime_1} = -6,\; \boldsymbol{e^\prime_2} = -1,\; \boldsymbol{e^\prime_3} = 2,\; \boldsymbol{e^\prime_4} = -8,\; \boldsymbol{e^\prime_5} = 6,\; \boldsymbol{e^\prime_6} = 0,\; \boldsymbol{e^\prime_7} = 0,\; \boldsymbol{e^\prime_8} = 6,\; \boldsymbol{e^\prime_9} = 0,\; \boldsymbol{e^\prime_{10}} =0 \big] \nonumber\\
    = \frac{1}{72}& \big( -6 \left[{\cal{S}}x^{}_a ( 3x^{}_b + D_{1b} ) \left( Q_0 - 2 \right) \bar\partial\cdot\bar\partial\cdot{\cal{K}}^0 \right] -\left[{\cal{S}}D^{}_{1a} ( 3x^{}_b + D_{1b} ) \left( Q_0 - 2 \right) \bar\partial\cdot\bar\partial\cdot{\cal{K}}^0 \right] \nonumber\\
    &\;\,+ 2 \left[{\cal{S}} ( 3x^{}_a + D_{1a} ) \left( Q_0 - 2 \right) ( 3x^{}_b + D_{1b} )\; \bar\partial\cdot\bar\partial\cdot{\cal{K}}^0 \right] - 8 \left[{\cal{S}} ( 3x^{}_a + D_{1a} ) \left( Q_0 - 2 \right) x^{}_b \bar\partial\cdot\bar\partial\cdot{\cal{K}}^0 \right] \nonumber\\
    &\;\, + 6 \left[{\cal{S}} Q_0 6 x^{}_a \bar\partial\cdot{\cal{K}}^0_{\cdot b} \right] + 6 \left[{\cal{S}} Q_0 x^{}_a D_{1b} \, \bar\partial\cdot\bar\partial\cdot{\cal{K}}^0 \right] \big)\,.
\end{align}
After performing straightforward calculations (employing Eqs. \eqref{xxxx} and \eqref{dddd}), we can simplify the result as shown below:
\begin{align}\label{ttGsolu0}
    Q_0 {\cal{K}}^{0}_{ab} =&\, - 2{\cal{S}} x^{}_a \bar\partial\cdot{\cal{K}}^0_{\cdot b} - D_{2a} \bar\partial\cdot{\cal{K}}^0_{\cdot b} + \frac{1}{72} \left( 32 R^2 \theta_{ab} - 4D_{2a} D_{1b} \right)\bar\partial\cdot\bar\partial\cdot{\cal{K}}^0_{} \nonumber\\
    &\, + \frac{1}{72} \left( D_{2a} D_{1b} - 2R^2\theta_{ab} \right) \left( Q_0 - 2 \right) \bar\partial\cdot\bar\partial\cdot{\cal{K}}^0_{} \,,
\end{align}
or alternatively, utilizing Eq. \eqref{Q1ten}, with Eq. \eqref{ttGsolu2}. Note that, by multiplying the above result by $x^a_{}$ and/or $\eta^{ab}_{}$, one can confirm its consistency. 
\end{widetext}
\end{appendix}


\end{document}